\def\year{2020}\relax
\documentclass[letterpaper]{article}
\usepackage{aaai20}
\usepackage{times}
\usepackage{helvet}
\usepackage{courier}
\usepackage{url}
\usepackage{graphicx}
\newcommand{\WidthPicLongTable}{0.24 }

\usepackage{booktabs}
\usepackage{amssymb}
\usepackage{amsmath}
\usepackage{mathabx}

\usepackage{times}
\usepackage{soul,color}
\usepackage{url}
\usepackage[hidelinks]{hyperref}
\usepackage[utf8]{inputenc}
\usepackage[small]{caption}
\usepackage{graphicx}
\usepackage{algorithm}
\usepackage{algorithmic}
\usepackage[utf8]{inputenc} 
\usepackage[T1]{fontenc}    
\usepackage{hyperref}       
\usepackage{url}            
\usepackage{amsfonts}       
\usepackage{nicefrac}       
\usepackage{microtype}      
\usepackage{xcolor}
\usepackage{graphicx}
\usepackage{multirow,array}
\usepackage{longtable}
\usepackage{pdflscape}
\usepackage{subcaption}

\title{Arena: A General Evaluation Platform and Building Toolkit for Multi-Agent Intelligence}

\author{
Yuhang Song,\textsuperscript{\rm 1} 
Andrzej Wojcicki,\textsuperscript{\rm 3} 
Thomas Lukasiewicz,\textsuperscript{\rm 1} 
Jianyi Wang,\textsuperscript{\rm 4} 
Abi Aryan,\textsuperscript{\rm 5} \\ \Large 
\textbf{Zhenghua Xu},\textsuperscript{\rm 1,2}\thanks{Corresponding author: Zhenghua Xu.} 
\textbf{Mai Xu},\textsuperscript{\rm 4} 
\textbf{Zihan Ding},\textsuperscript{\rm 6} 
\textbf{Lianlong Wu}\textsuperscript{\rm 1}\\ 
\textsuperscript{\rm 1}Department of Computer Science, University of Oxford, United Kingdom\\
\textsuperscript{\rm 2}State Key Laboratory of Reliability and Intelligence of Electrical Equipment, Hebei University of Technology, China\\
\textsuperscript{\rm 3}Lighthouse, 
\textsuperscript{\rm 4}School of Electronic and Information Engineering, Beihang University, China\\
\textsuperscript{\rm 5}University of California, Los Angeles, United States, 
\textsuperscript{\rm 6}Imperial College London, United Kingdom\\
\{yuhang.song,thomas.lukasiewicz\}@cs.ox.ac.uk, andrzej@wojcicki.xyz, zhenghua.xu@hebut.edu.cn
}

\begin{document}
\maketitle
\begin{abstract}
Learning agents that are not only capable of taking tests, but also innovating is becoming a hot topic in AI. One of the most promising paths towards this vision is multi-agent learning, where agents act as the environment for each other, and improving each agent means proposing new problems for others. However, existing evaluation platforms are either not compatible with multi-agent settings, or limited to a specific game. That is, there is not yet a general evaluation platform for research on multi-agent intelligence. To this end, we introduce Arena, a general evaluation platform for multi-agent intelligence with 35 games of diverse logics and representations. Furthermore, multi-agent intelligence is still at the stage where many problems remain unexplored. Therefore, we provide a building toolkit for researchers to easily invent and build novel multi-agent problems from the provided game set based on a GUI-configurable social tree and five basic multi-agent reward schemes. Finally, we provide Python implementations of five state-of-the-art deep multi-agent reinforcement learning baselines. Along with the baseline implementations, we release a set of 100 best agents/teams that we can train with different training schemes for each game, as the base for evaluating agents with population performance. As such, the research community can perform comparisons under a stable and uniform standard. All the implementations and accompanied tutorials have been open-sourced for the community at https://sites.google.com/view/arena-unity/.
\end{abstract}

\section{Introduction}
\label{sec-introduction}

Modern learning algorithms are more of outstanding test-takers, but less of innovators, i.e., the ceiling of an agent's intelligence may be limited by the complexity of its environment \cite{leibo2019autocurricula}.
Thus, the emergence of innovation is becoming a hot topic for AI.
One of the most promising paths towards such a vision is learning via social interaction, i.e., multi-agent learning.
In multi-agent learning, how the agents should beat the opponents or collaborate with each other is not defined or limited by the creator of the environment, e.g., the inventor of the ancient Go never defines what strategies are good.
However, enormous and sophisticated strategies are invented while a population of human players/artificial agents evolves by improving themselves over the others, i.e., each agent is acting as an environment for the others and improving itself means proposing new problems for the others.

\begin{figure*}
	\centering
	\includegraphics[width=\textwidth]{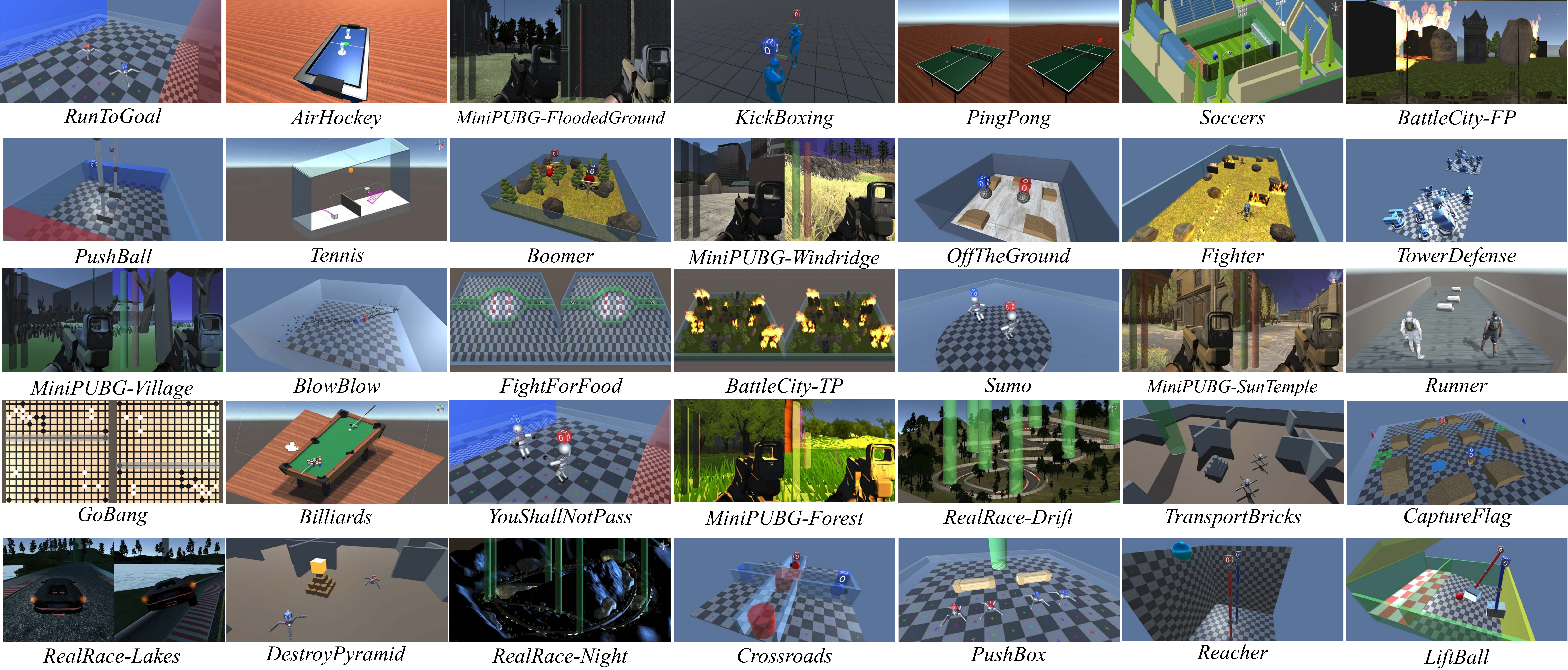}
	\caption{Game set of Arena.}
	\label{game-set-arena}
\end{figure*}

To study a new class of intelligence, general evaluation platforms with diverse games are 
milestones that push forward the research to the next levels. For example, \textrm{ALE} \cite{bellemare2013arcade}, \textrm{Mujoco} \cite{todorov2012mujoco}, and \textrm{DM-Suite} \cite{tassa2018deepmind} are the most spread general evaluation platforms that greatly accelerate the research in general reinforcement learning.
However, there is no such general evaluation platform for multi-agent intelligence.
Although some platforms support multi-agent settings \cite{wydmuch2018vizdoom,vinyals2017starcraft}, they are not general evaluation platforms, i.e., built for specific games.
Thus, in this paper, we propose the first general evaluation platform for multi-agent intelligence, called \textit{Arena}, containing 35 multi-agent games in total, with diverse logics and representations; see Fig.\ \ref{game-set-arena}.

\begin{figure*}
	\centering
	\includegraphics[width=0.8\textwidth]{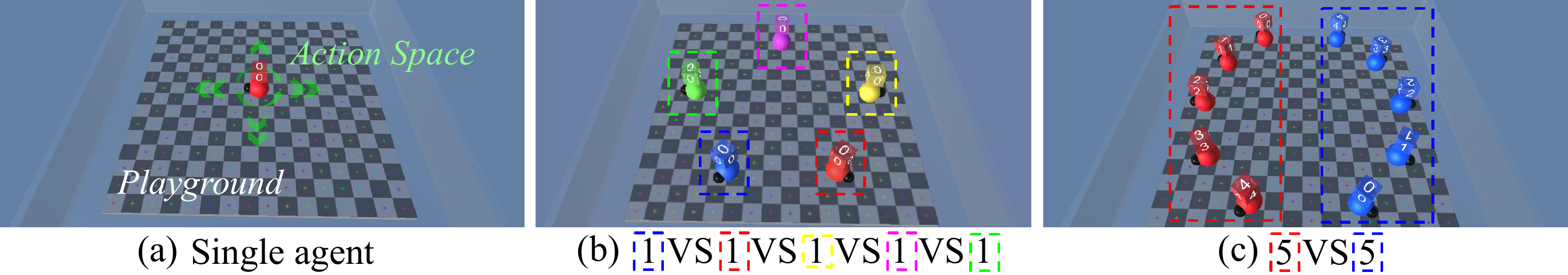}
	\caption{Game examples of the extensible multi-agent platform.}
	\label{customize-example}
\end{figure*}

Apart from training and evaluation, multi-agent intelligence research is still at a stage where many problems remain undiscovered or unexplored.
Thus, the second contribution of Arena is a building toolkit for multi-agent intelligence, enabling the easy creation of different multi-agent scenarios.
For example, in the sample game in Fig.\ \ref{customize-example} (a), after defining the basic behavior of the agent (i.e., moving and turning) and the ``alive'' state of the agent (i.e., it stays on the playground), it can be extended to different multi-agent scenarios with minimal effort.
For example, (1) five players fight each other until only one agent is left alive (see Fig.\ \ref{customize-example} (b)), or (2) $5 \times 2$ players form 2 teams, and each agent fights for its own team until all players in a team are dead (see Fig.\ \ref{customize-example} (c)), or (3) multiple players form multiple teams in hierarchies, where the collaboration and competition relationships between the teams are customized (see Fig.\ \ref{example-tree}).

Thus, Arena is not just a research platform for the evaluation with a fixed set of games, but also a building toolkit for researchers to invent and build novel multi-agent problems.

To achieve the above vision of building a toolkit for multi-agent intelligence, (1) we provide a GUI-configurable tree that defines the social structure of agents, called \textit{social tree}; and (2) based on the social tree, we propose 5 basic multi-agent reward schemes (BMaRSs),  which define different social paradigms at each node in the social tree.
Specifically, each BMaRS is a restriction applied to the reward function, so it corresponds to a batch of reward functions that can lead to a specific social paradigm.
For each BMaRS, Arena provides multiple ready-to-use reward functions, simplifying the construction of games with complex social relationships.
Furthermore, if the agent is controlling each joint of a robot, it has long been a burden for researchers that low-level intelligence (such as the basic skill of moving) must first be built, before they can study high-level multi-agent intelligence \cite{heess2017emergence}.
Thus, Arena provides many ready-to-use dense reward functions in each BMaRS that handle such low-level intelligence.
Additionally, Arena also offers a verification option for customized reward functions, so the researchers can make sure that the programmed reward functions lie in one of the BMaRSs that produces a specific social paradigm.
Thus, with the above efforts towards a building toolkit for multi-agent intelligence and the provided set of 35 games for a general evaluation platform, one can easily customize a set of games of a new social paradigm to study a yet unexplored problem.

Finally, we provide Python implementations of several state-of-the-art deep multi-agent reinforcement learning baselines, which can be used as starting points for the development of novel multi-agent algorithms, as well as the validation of new environments.
Along with the baseline implementations, we also release a set of 100 best agents/teams that we can train with different training schemes for each game, as the base for evaluating agents with population performance \cite{balduzzi2019open,balduzzi2018re}.
So, the research community can perform comparisons under a stable and uniform standard.

To summarize, this paper's contributions are as follows:
(1) a general evaluation platform for multi-agent intelligence with a set of diverse games, most of which are new to the community or still stand as a challenge for state-of-the-art algorithms,
(2) a building toolkit for multi-agent games, enabling the easy creation of new social paradigms based on GUI-configurable social trees and BMaRSs,
(3) the baseline implementations of 5 state-of-the-art multi-agent algorithms for both competitive and collaborative settings, and
(4) sets of benchmark agents/teams for the community to conduct stable and uniform population evaluation \cite{balduzzi2018re}.
Code for games, building toolkit, and baselines, as well as all corresponding tutorials have been released online at https://sites.google.com/view/arena-unity/.

\section{The Platform}

\textbf{State-of-the-Art Engine.} The engine behind Arena is the world-leading game engine \textit{Unity} \cite{juliani2018unity}, which provides Arena with several desirable features on rendering, physics, customizability, and community.
There are also other choices of popular engines.
Some platforms contain a wide set of diverse games \cite{bellemare2013arcade,nichol2018gotta,perez2016general,OpenAI_Universe}. 
However, they are designed mostly for single-agent scenarios and are extremely hard to customize (adding multiple players or creating new games), since the games are provided as compiled binary ROMs.
Other downsides of these choices include deterministic environments, unrealistic rendering, and unrealistic physics.
Other platforms \cite{todorov2012mujoco,tassa2018deepmind} are, in nature, more physics engines than game engines, which lack a visual editor 
for easily creating customized games, and cannot handle more ``game-like'' features, such as instantiating and destroying objects in real-time during the simulation.
The rest of the platforms are limited in the sense that they are built for specific tasks, such as for first-person shooting \cite{wydmuch2018vizdoom}, Real-Time Strategy (RTS) \cite{tian2017elf}, vision understanding \cite{qiu2017unrealcv}, in-door scene understanding \cite{handa2016scenenet,brodeur2017home,savva2017minos,chang2017matterport3d,puig2018virtualhome,gao2019vrkitchen}, surviving \cite{suarez2019neural}, and interaction \cite{wu2018building,savva2019habitat,kolve2017ai2}, or specific games, such as Starcraft \cite{vinyals2017starcraft} and Dota2 \cite{openai2018dota}.
Thus, creating a general evaluation platform on these engines is not a reasonable choice.
\textrm{DeepMind Lab} \cite{beattie2016deepmind}, \textrm{Psychlab} \cite{leibo2018psychlab}, and \textrm{Malmo} \cite{johnson2016malmo} are more appropriate choices when building a customizable general evaluation platform.
However, the main drawbacks of the above engines are tied to their dated nature.
The rendering system of these engines are either low-polygon pixelated (Malmo, based on Minecraft) or outdated (DeepMind Lab and Psychlab, based on Quake III).
The physics systems of these engines are either rudimentary (Malmo), or have a gap \cite{juliani2018unity} to the physical world (DeepMind Lab and Psychlab).
Besides, they are all incompatible with a visual editor, making it quite cumbersome to build customized scenarios.

To summarize, built on Unity, Arena has the following advantages over other platforms:
(1) realistic rendering, so that features, such as complex lighting, textures, and shaders, are fully handled by the background engine and easily produced in a customized game,
(2) realistic physics, so that enough and realistic stochasticity is introduced in the game and transferring a policy learned within a simulator to the real world is easier,
(3) user-friendly visual editor, so that building new multi-agent scenarios in Arena is easy,
and (4) a large and active development community, so that creating new games is easy with millions of off-the-shelf assets.

\smallskip\noindent\textbf{Game Sets Towards General Intelligence.} The first contribution of Arena is to provide a set of multi-agent games with diverse game logics and representations, so that it may push forward the research of general multi-agent intelligence.
Specifically, Arena provides:
(1) 27 new games that are not yet studied in the community,
(2) 8 games, of which the basic logics are inspired by other research, but equipped with realistic rendering effects, physics engine, and all features described in the following two paragraphs, such as extensibility to other social paradigms,
and (3) interface to the popular stand-alone domain StarCraft.
The game set is shown in Fig. \ref{game-set-arena}.
For more detailed information, see Tables 2-7 in the extended paper \cite{supplementary_material}.

\smallskip\noindent\textbf{Building Toolkit for Multi-Agent Environments.}
As the second contribution, we provide a building toolkit for multi-agent environments: we provide (1) a GUI-configurable social tree that defines how agents are grouped together with each other, and (2) 5 basic multi-agent reward schemes (BMaRSs) applied on each node in the social tree, so that different social relationships can be easily built and verified, and low-level intelligence (like motor skills) can be handled.

\smallskip\noindent\textbf{Other Features.} Learning to communicate is an important research area in multi-agent intelligence \cite{das2017learning,mordatch2018emergence}.
Thus, Arena provides a broadcast board at each node of the social tree (accessible for any agent as a child of the node), which enables the study of learning communication at each level.
Also, global states may be used in research for different purposes \cite{lowe2017multi,gupta2017cooperative,foerster2017stabilising,foerster2018counterfactual}.
Thus, Arena provides the option to broadcast it to all agents.
Besides, a top-down view of the global game is often appreciated for visualizing population behavior \cite{johnson2016malmo,wydmuch2018vizdoom,jaderberg2018human,liu2019emergent}.
Thus, Arena by default enables this option.
Finally, there is a necessity for competitive agents to evaluate against human players, and also a research trend for collaborative agents to team up with human players.
Thus, Arena provides a gaming interface for humans, so that a human player can take the place of any agent in the game.

\section{Basic Multi-Agent Reward Schemes and Social Trees}
\label{sec-basic-reward-schemes}

\textbf{Preliminaries.}
We consider a Markov game as defined in \cite{littman1994markov},
consisting of multiple agents $x \in \mathcal{X}$,
a finite global state space $\mathcal{S}$,
a finite action space $\mathcal{A}_{x}$ for each agent $x$, and a bounded-step reward space $r_{x,t} \in \mathbb{R}$ for each agent $x$.
The environment consists of
a transition function $g: \mathcal{S}\times \bigtimes \bigl\{ \mathcal{A}_{x} : x \in \mathcal{X}  \bigl\} \rightarrow \mathcal{S} $, which is a stochastic function $s_{t+1} \sim g \bigl ( s_{t},  (a_{x,t})_{x \in \mathcal{X}} \bigl )$,
a reward function  for each agent $f_{x}: \mathcal{S} \times\bigtimes \bigl\{ \mathcal{A}_{x} : x \in \mathcal{X} \bigl\} \rightarrow \mathbb{R} $, which is a deterministic function $r_{x,t+1} = f_{x} \bigl ( s_{t}, (a_{x,t})_{x \in \mathcal{X}} \bigl )$, a joint reward function $f= ( f_{x} )_{x \in \mathcal{X}}$,
and episode reward $R^{f}_{x} = \sum_{t=1}^{T} r_{x,t}$ for each agent $x$ under the joint reward function $f$.
For the agent, we consider that it observes $s_{x,t} \in \mathcal{S}_{x}$, where $\mathcal{S}_{x}$ consists of a part of the information from the global state space $\mathcal{S}$.
Thus, we have a policy $\pi_{x}: \mathcal{S}_{x} \rightarrow \mathcal{A}_{x} $, which is a stochastic function $a_{x,t} \sim \pi_{x} (s_{x,t}) $.
Besides, we consider that the agent $x$ can take a policy $\pi_{x}$ from a set of policies $\Pi_{x}$ and assume that the random seed of all sampling operations is $k$, which is sampled from the whole seed space~$\mathcal{K}$.

We investigate the effect of $\{ x : x \in \mathcal{X} \}$ and $\{ \pi_{x} : \pi_{x} \in \Pi_{x} \}$ on $\{ R^{f}_{x} : x \in \mathcal{X} \}$.
By applying different restrictions on the effect, we have different BMaRSs, each one of which is a set of joint reward functions $\mathcal{F} = \{ f : \cdot \} $ that produce a similar effect on the population $\mathcal{X}$.
The term reward scheme first appears in \cite{tampuu2017multiagent} as a tabular, which is applied to a special case of Pong.
While we define it in a general form and show that many examples are special cases within this general form.

In a non-sequential setting (normal-form game), the reward scheme serves a similar purpose as the payoff matrix \cite{myerson2013game}, which is also represented as a tabular.
See Lemma 2 in the extended paper \cite{supplementary_material} for how the payoff matrix is aligned with BMaRSs.
In the following, we define 5 different BMaRSs.
Along defining these BMaRSs, we also describe the ready-to-use reward functions $f$ within these BMaRSs, which is provided by Arena as a dropdown list.

\smallskip\noindent\textbf{Non-learnable} BMaRSs ($\mathcal{F}^{NL}$) are a set of joint reward functions $f$ as follows:
\begin{equation}
\begin{split}
\label{non-learnable}
\mathcal{F}^{NL} = \bigl\{
	f :
	\forall k \in \mathcal{K},
	\forall x \in \mathcal{X},
	\forall \pi_{x} \in \Pi_{x}, \\
	{\partial R^{f}_{x}}\,/\,{\partial \pi_{x}} = \mathbf{0}
\bigr\}, 
\end{split}
\end{equation}
where $\mathbf{0}$ is a zero matrix of the same size and shape as the parameter space that defines $\pi_{x}$.
Intuitively, $\mathcal{F}^{NL}$ means that $R^{f}_{x}$ for any agent $x \in \mathcal{X}$ cannot be optimized by improving its policy $\pi_{x}$.

\smallskip\noindent\textbf{Isolated} BMaRSs ($\mathcal{F}^{IS}$) are a set of joint reward functions $f$ as follows:

\begin{equation}
\begin{split}
\label{isolated}
\mathcal{F}^{IS} = \bigl\{
	f :
		f \notin \mathcal{F}^{NL}
	\textrm{~and~}
		\forall k \in \mathcal{K},
		\forall x \in \mathcal{X}, \\
		\forall x' \in \mathcal{X} \setminus \{x\},
		\forall \pi_{x} \in \Pi_{x},
		\forall \pi_{x'} \in \Pi_{x'},
		\frac{\partial R^{f}_{x}}{\partial \pi_{x'}} = \mathbf{0}
		\bigr\}, 
\end{split}
\end{equation}
Intuitively, $\mathcal{F}^{IS}$ means that the episode reward $R^{f}_{x}$ received by any agent $x \in \mathcal{X}$ is not related to any policy $\pi_{x'}$ taken by any other agent $x' \in \mathcal{X} \setminus \{x\}$.

Reward functions $f_{x}$ in $f$ of $\mathcal{F}^{IS}$ are often called \textit{internal reward functions} in other multi-agent approaches \cite{hendtlass2004introduction,jaderberg2018human,bansal2018emergent}, meaning that apart from the reward functions applied at a population level (such as win/loss), which are too sparse to learn, there are also reward functions directing the learning process towards receiving the population-level rewards, but are more frequently available, i.e., more dense \cite{heess2017emergence}.
$\mathcal{F}^{IS}$ is especially practical if the agent is a robot requiring continuous control of applying force on each of its joints, which means basic motor skills (such as moving) need to be learned before generating population-level intelligence.
Thus, we provide $f$ in $\mathcal{F}^{IS}$ of energy cost, punishment of applying a big force, encouragement of keeping a steady velocity, and moving distance towards target.

\smallskip\noindent\textbf{Competitive} BMaRSs ($\mathcal{F}^{CP}$), inspired by \cite{cai2011minmax}, are defined as
\begin{equation}
\begin{split}
\label{competitive-eq}
\mathcal{F}^{CP} = \bigl\{
	f :
		f \notin \mathcal{F}^{NL} \cup \mathcal{F}^{IS}
	\textrm{~and~}
		\forall k \in \mathcal{K},
		\forall x \in \mathcal{X}, \\
		\forall \pi_{x} \in \Pi_{x},
		\forall \pi_{x'} \in \Pi_{x'},
		\frac{\partial \int_{x' \in \mathcal{X}} R^{f}_{x'} dx'}{\partial \pi_{x}} = \mathbf{0}
\bigr\}, 
\end{split}
\end{equation}
which intuitively means that for any agent $x \in \mathcal{X}$, taking any possible policy $\pi_{x} \in \Pi_{x}$, the sum of the episode reward of all agents will not change.
If the episode length is $1$, it expresses a classic multi-player zero-sum game \cite{cai2011minmax}.
Useful examples of $f$ within $\mathcal{F}^{CP}$ are: (1)~agents fight for a limited amount of resources that are always exhausted at the end of the episode, and the agent is rewarded for the amount of resources that it gained, and (2) fight till death, and the reward is given based on the order of death (the reward can also be based on the reversed order, so that the one departing the game first receives the highest reward, such as in some poker games, the one who first discards all cards wins).
\textit{Rock, Paper, and Scissors} in normal-form game \cite{myerson2013game} and \textit{Cyclic Game} in \cite{balduzzi2019open} are both special cases of $\mathcal{F}^{CP}$; see Lemmas 2 and 3 in the extended paper \cite{supplementary_material}.

\smallskip\noindent\textbf{Collaborative} BMaRSs ($\mathcal{F}^{CL}$), inspired by \cite{cai2011minmax}, are defined as
\begingroup
\begin{equation}
\begin{split}
\label{collaborative}
\mathcal{F}^{CL} = \bigl\{
	f :
		f \notin \mathcal{F}^{NL} \cup \mathcal{F}^{IS}
	\textrm{~and~}
		\forall k \in \mathcal{K},
		\forall x \in \mathcal{X}, \\
		\forall x' \in \mathcal{X} \setminus \{x\},
		\forall \pi_{x} \in \Pi_{x},
		\forall \pi_{x'} \in \Pi_{x'},
	\frac{\partial R^{f}_{x'}}{\partial R^{f}_{x}} \geq 0
\bigr\}, 
\end{split}
\end{equation}
\endgroup
which, intuitively, means that there is no conflict of interest (${\partial R^{f}_{x'} }\,/\,{\partial R^{f}_{x} } < 0$) for any pair of agents $(x', x)$.
Besides, since $f \notin \mathcal{F}^{NL} \cup \mathcal{F}^{IS}$, there is at least one pair of agents $(x, x')$ that makes ${\partial R^{f}_{x'} }\,/\,{\partial R^{f}_{x}} > 0$.
This indicates that this pair of agents shares a common interest, so that improving $R^{f}_{x}$ for agent $x$ means improving $R^{f}_{x'}$ for agent $x'$.
The most common example of $f$ within $\mathcal{F}^{CL}$ is that $f_{x}$ for all $x \in \mathcal{X}$ is identical, such as the moving distance of an object that can be pushed forward by the joint effort of multiple agents, or the alive duration of the population (as long as there is at least one agent alive in the population, the population is alive).
Thus, we provide $f$ in $\mathcal{F}^{CL}$:  living time of the team (both positive and negative, since some games require the team to survive as long as possible, while other games require the team to depart as early as possible, such as poker).

\smallskip\noindent\textbf{Competitive and Collaborative Mixed} BMaRSs ($\mathcal{F}^{CC}$) are defined as a catch-all for any other than the above four ones.
First, the term ${\partial \int_{x' \in \mathcal{X}} R^{f}_{x'} dx'}\,/\,{\partial \pi_{x}} = \mathbf{0}$ in \eqref{competitive-eq}
can be written as $\int_{x' \in \mathcal{X} } { \partial R^{f}_{x'}}\,/\,{\partial R^{f}_{x}} dx' = 0$ (see Lemma 1 in
extended paper \cite{supplementary_material}, which makes an alternative \eqref{competitive-eq}.
Considering  $\mathcal{F}^{CP}$ in this alternative \eqref{competitive-eq} and $\mathcal{F}^{CL}$ in \eqref{collaborative}, an intuitive explanation of $\mathcal{F}^{CC}$ is that there exist circumstances when ${\partial R^{f}_{x'}}\,/\,{\partial R^{f}_{x}} < 0$, meaning that the agents are competitive at this point.
But the derivative of total interest  $\int_{x' \in \mathcal{X} }
{ \partial R^{f}_{x'}}\,/\,{\partial R^{f}_{x}} dx'$ is not always $0$; thus, the total interest can be maximized with specific policies, meaning that the agents are collaborative at this point.

Apart from providing several practical $f$ in each BMaRS, we also provide a verification option for each BMaRS, meaning that one can customize an $f$ and use this verification option to make sure that the programmed $f$ lies in a specific BMaRS.
The implementation of verification option can be found in Section 1 in the extended paper \cite{supplementary_material}.

\begin{figure*}
	\centering
	\centerline{\includegraphics[width=0.8\textwidth]{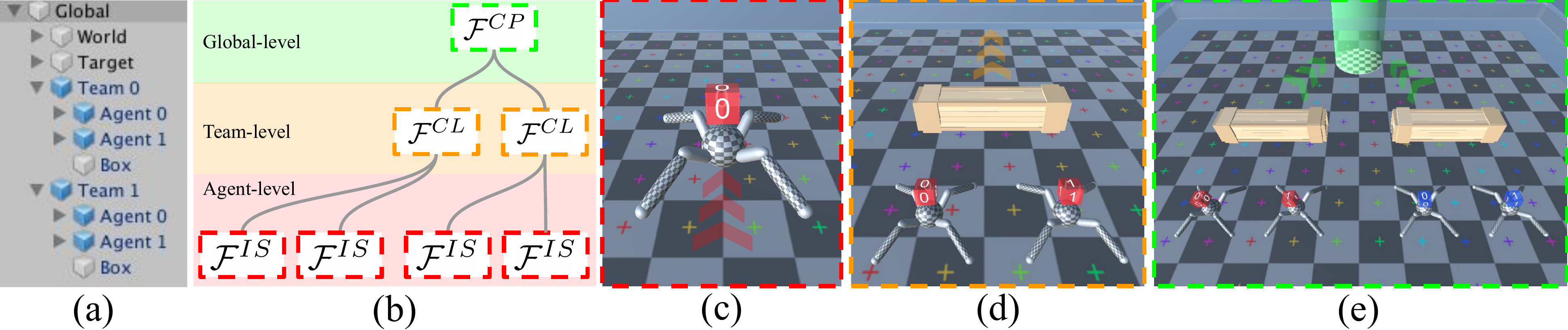}}
	\caption{An example of a social tree and BMaRSs applied on it.}
	\label{example-tree}
\end{figure*}

\smallskip\noindent\textbf{The Social Tree.} The BMaRSs defined above apply to an agent group of all sizes.
To define more complex and structured social paradigms, we use a tree structure (\textit{social tree}) to organize the agents and apply BMaRSs on each node of the tree.
We illustrate this by an example.
The GUI interface in Fig.\ \ref{example-tree} (a) defines a tree structure in Fig. \ref{example-tree} (b), representing a population of 4 agents.

The tree structure can be easily reconfigured by dragging, duplicating, or deleting nodes in the GUI interface in Fig.\ \ref{example-tree} (a).
In this example, each agent has an agent-level BMaRS.
The agent is a robot ant, so that the agent-level BMaRSs are $\mathcal{F}^{IS}$, specifically, the option of \textit{ant-motion} that directs the learning towards basic motion skills such as moving forward, as shown in Fig.\ \ref{example-tree} (c).
Each two agents form a \textit{team} (which is a set of agents or teams), the two agents have team-level BMaRSs.
In this example, the two robot ants collaborate with each other to push a box forward, as shown in Fig.\ \ref{example-tree} (d).
Thus, the team-level BMaRSs are $\mathcal{F}^{CL}$, specifically, the moving distance of the box.
On the two teams, we have global-level BMaRSs.
In this example, the two teams are set to have a match regarding which team pushes its box to the target point first, as shown in Fig.\ \ref{example-tree}~(e).
Thus, the global-level BMaRSs are $\mathcal{F}^{CP}$, specifically, the ranking of the box reaching the target.
The final reward function applied to each agent is a weighted sum of the above three BMaRSs at three levels.
One can imagine defining a social tree of more than three levels, where small teams form bigger teams, and BMaRSs are defined at each node to give more complex and structured social problems.
After defining the social tree and applying BMaRSs on each node, the environment is ready for use with an abstraction layer handling everything else, such as assigning viewports to each agent in the window, applying the team color, displaying the agent ID, and generating a top-down view.

\section{The Learning Agents}
\label{sec-learning-agents}

\textbf{The Baselines.}
We provide Python implementations of several state-of-the-art baselines that can be used as starting points for the development of novel multi-agent algorithms, as well as for the validation of new environments.
Specifically, we first implement a fully decentralized system, where each agent is a self-contained PPO \cite{schulman2017proximal}, with independent critic, actor, and optimizer. We also implement two state-of-the-art methods based on self-play in \cite{openai2018dota} (SP) and population-based training in
\cite{jaderberg2018human,AlphaStar} (PB).
For collaborative agents, we implement two state of the arts: centralized critic \cite{lowe2017multi} (CC) and centralized critic with a counterfactual baseline \cite{foerster2018counterfactual} (CF).

\smallskip\noindent\textbf{The Evaluation Metric.}
It is recently raising attention that evaluating an agent against a single-agent or hand-coded bot is unstable and misleading \cite{balduzzi2018re}.
Thus, the population performance is introduced to evaluate an agent's (or an agent group's) performance among a base population.
To enable population evaluation, we release 100 best agents,  which we can train with different training schemes for each game as the base population.
One can call the provided function to get the ranking of an agent among the base population, or get the averaged ranking of a population among the base population.
Moreover, we provide a human ranking among the base population, which provides an indication of human-level intelligence in the game.
We will accept the submission of agents from the community as well as keep implementing algorithms introduced in the future, so that the base population will be upgraded, as the level of research in multi-agent intelligence advances.

\section{Experiments}
\label{sec-experiments}

Experiments are conducted from three aspects.
First, we evaluate our game set from the perspective of stochasticity, realistic rendering, and simulation speed, all of which are presented in the extended paper \cite{supplementary_material} due to page limit.
Other advantages from the Unity engine have been verified by \cite{juliani2018unity}.
Second, we evaluate our design of the extensible multi-agent building toolkit with a case study, showing that by applying different social trees and BMaRSs, different population-level strategies can be learned.
Third, we report the experimental results of 5 baselines that we implemented and show that by using the provided population performance evaluation metric, the training progress can be visualized in a less noisy and more analyzable way.

\begin{figure*}
	\centering
	\centerline{\includegraphics[width=0.85\textwidth]{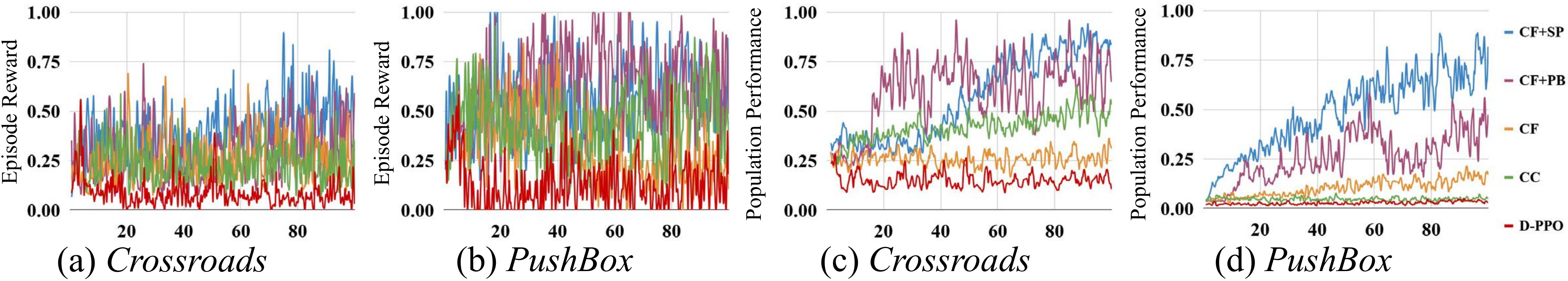}}
	\caption{Visualizing training progress over episode reward (a,b) and population performance (c,d) of different baselines: D-PPO (Decentralized Proximal Policy Optimization), SP (Self-Play), PB (Population-Based training), CC (Centralized Critic), and CF (Counterfactual Baseline).}
	\label{baselines}
\end{figure*}

\begin{figure}
	\centering
	\includegraphics[width=0.8\columnwidth]{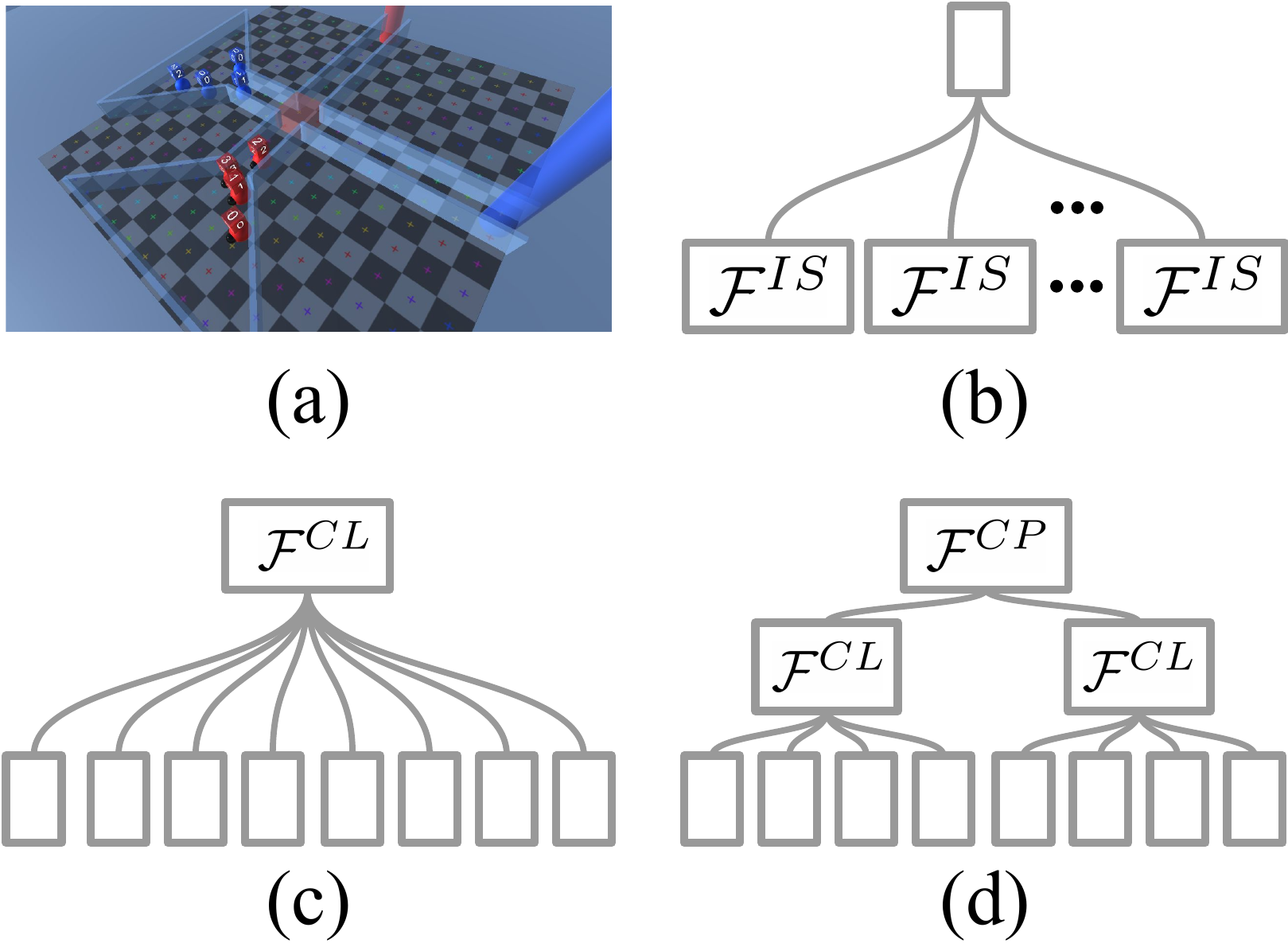}
	\caption{Case Study of Social Tree and BMaRSs.}
	\label{crossroads}
\end{figure}

\smallskip\noindent\textbf{Case Study of Social Tree and BMaRSs.}
We use the game \textit{Crossroads} from Arena to study the effectiveness of the proposed social tree and BMaRSs via designing different social paradigms.
Specifically, in the game \textit{Crossroads} shown in Fig. \ref{crossroads} (a), the agent can move and turn, the final goal of the agent is to reach the target on the other side of the crossroad.
By defining different social trees and applying different BMaRSs, as shown in Fig. \ref{crossroads} (b) to (d), the agents learn different strategies.
In Fig. \ref{crossroads} (b), \textit{isolated} BMaRSs ($\mathcal{F}^{IS}$) are applied to all agents, i.e., each agent minimizes the time that it takes to reach the target.
The result shows that the learned agents simply rush forward,  and they easily crash with each other at the center of the crossroad, producing a traffic jam.
In Fig. \ref{crossroads} (c), \textit{collaborative} BMaRSs ($\mathcal{F}^{CL}$) are applied to the parent node of all agents, i.e., all agents are rewarded with the time that the last one of them takes to reach the target.
The result shows that the agents learn to wait for each other to go across the crossroad, so that they can all get across as efficiently as possible.
In Fig. \ref{crossroads} (d), \textit{collaborative}  BMaRSs ($\mathcal{F}^{CL}$) are applied on the parent node of every 4 agents (which form a team), and  \textit{competitive} BMaRSs ($\mathcal{F}^{CP}$) are applied on the parent node of the two teams.
Specifically speaking, each two agents in the same team are rewarded with the same reward, and the reward is $1$ for the team that gets all of its agents to the target first, $0$ for the other team.
The results show that each team learns to block the road of the other team with one agent, so that the other agents in the team can get across undisturbed.
Then, the agent that blocks the road leaves for the target,  after all its teammates have reached the target.

\smallskip\noindent\textbf{Baselines and Evaluation Metric.}
We compare 5 baselines on two games: (1) \textit{Crossroads} in Fig. \ref{crossroads} (a) with the BMaRS settings of Fig. \ref{crossroads} (d)  and (2) \textit{PushBox} in Fig. \ref{example-tree} (e) with the BMaRS settings of Fig. \ref{example-tree} (b).
The BMaRS settings of both games contain competitive as well as collaborative social relationships, i.e., multiple agents form collaborative teams, and teams compete with each other.
Thus, we investigate SP and PB baselines at the level of teams competing with each other, as well as investigate CC and CF baselines at the level of agents collaborating with each other in a team.
As can be seen, the curve of episode reward shown in Fig. \ref{baselines} (a) and (b) is extremely noisy, as the environment is non-stationary with the strategy of other collaborators and/or competitors evolving during the training.
However, in Fig. \ref{baselines} (c) and (d), which is the curve of ranking in the released base population, i.e., population performance, all methods are comparable with clear performance gaps.

\section{Related Work}
\label{sec-related-work}

Surveys of multi-agent intelligence research can be found in  \cite{hernandez2018multiagent}.
Different ideas have been explored on competitive and collaborative multi-agent settings.

\smallskip\noindent\textbf{Collaborative Settings.}
The simplest way to deploy multi-agent collaborative systems is to make each agent have a completely independent learning process (fully decentralized) \cite{matignon2012independent}.
However, collaborative behaviors are hardly observed under such fully decentralized setting; thus, a fully centralized system is utilized in \cite{peng2017multiagent}, where the policy has access to the global state and is shared by all agents.
However, it is impractical, since the global state is mostly unavailable in practice, and the system does not support extending the number of agents.
Thus, centralized training and decentralized execution are gaining attention \cite{kraemer2016multi}.
For multi-agent systems, this idea is mostly explored under actor-critic algorithms \cite{foerster2018counterfactual}.
Other ideas include using a joint action-value function, \cite{lauer2004reinforcement}
addressing the variance problem by a large batch size \cite{bansal2018emergent}, 
and learning grounded cooperative communication protocols between agents \cite{foerster2016learning}.

\smallskip\noindent\textbf{Competitive Settings.}
Competitive multi-agent intelligence  originally comes from computational game theory \cite{bowling2015heads}.
Later on, deep multi-agent reinforcement learning (D-MARL) is preferred, due to its scalability, and as it achieves notable advances on two-player games, such as Poker and Go \cite{moravvcik2017deepstack,silver2017mastering}.
Later, D-MARL was applied to more diverse problems, such as high-dimensional video games \cite{openai2018dota,AlphaStar} and those involving physics control \cite{bansal2018emergent}.
When solving more practical problems, many issues have been raised, such as ensuring diversity amongst agents \cite{marivate2015improved},
avoiding overfitting to the policy of the opponents \cite{lanctot2017unified}.
Many ideas address such issues \cite{kleiman2016coordinate}.
Following on D-MARL, a very promising recent direction  is self-play \cite{tesauro1995temporal}.
Fictitious self-play \cite{heinrich2016deep}
first shows promising performance on the competitive game Leduc Poker.
However, as the stability and parallelizability are improving with the invention of new reinforcement learning algorithms, state-of-the-art approaches adopt a simpler form of self-play \cite{openai2018dota}, which produces a superior-human intelligence on large video games, like Dota2.
Another promising recent idea  is population-based training, as adopted in StarCraft \cite{AlphaStar}.

\section{Summary and Outlook}

This paper has introduced the first general evaluation platform for multi-agent intelligence research.
Besides, with the efforts on a building toolkit of multi-agent environments, the platform also allows for easily building new multi-agent problems.
Additionally, with the released implementations of several state-of-the-art baselines, researchers can start their adventure instantly.
Finally, by releasing a base population, the community can conduct comparisons under a stable and uniform evaluation metric.

\subsubsection{Acknowledgments.}
This work was supported by the China Scholarship Council under the State Scholarship Fund, by the Graduate Travel and Special Project Grants from the Somerville College of the University of Oxford, by the Alan Turing Institute under the UK EPSRC grant EP/N510129/1, by the AXA Reseach Fund, by the National Natural Science Foundation of China under the grants 61906063, 61876013, and 61922009, by the Natural Science Foundation of Tianjin under the grant 19JCQNJC00400, by the ``100 Talents Plan’’ of Hebei Province, and by the Yuanguang Scholar Fund of Hebei University of Technology.

\clearpage
\onecolumn

\section{Experiments: Realistic Rendering, Simulation Speed and Stochasticity}

\begin{figure}
	\centering
	\includegraphics[width=0.6\columnwidth]{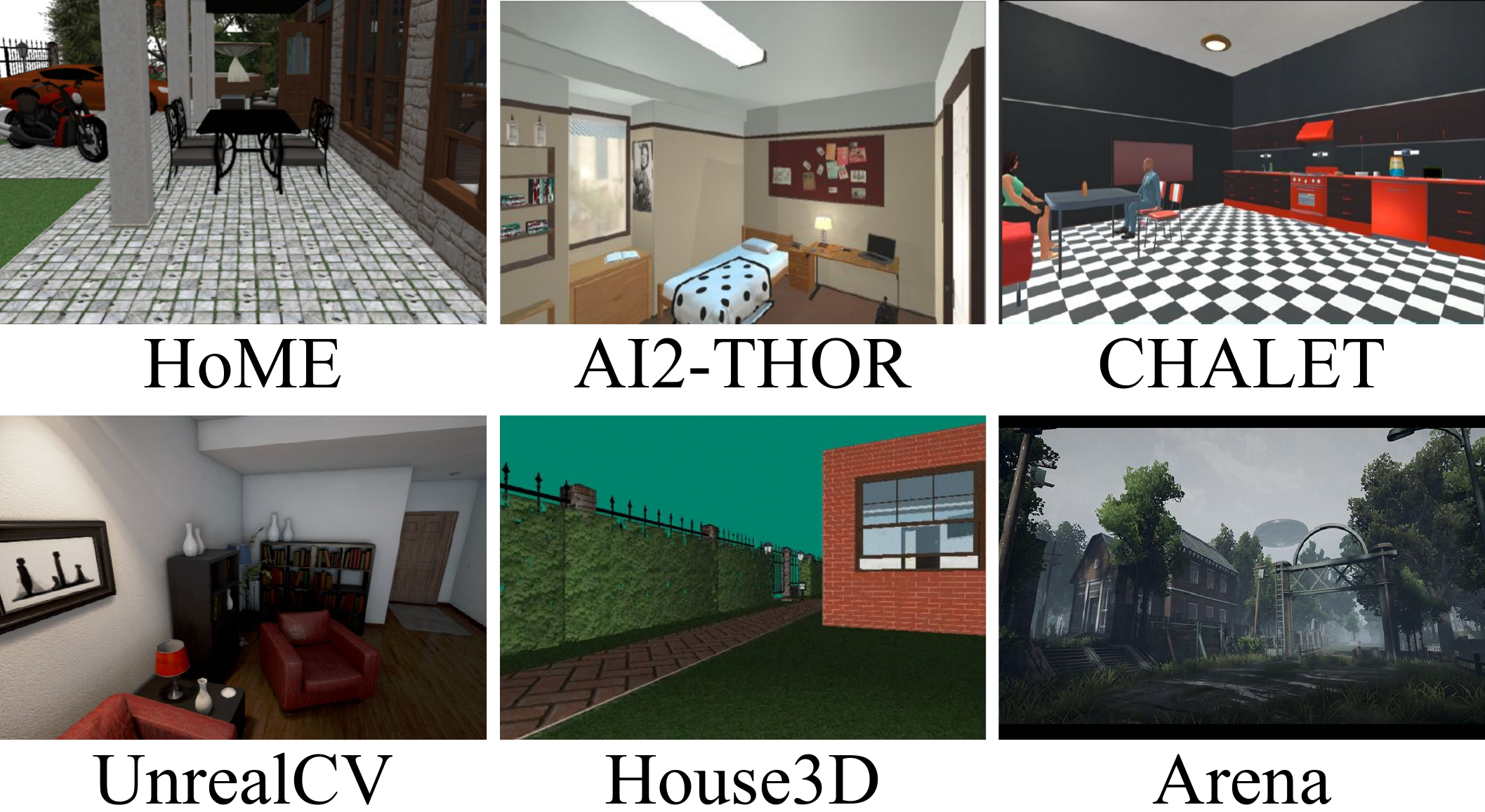}
	\caption{Comparison of realistic rendering effect.}
	\label{fig-realistic}
\end{figure}

\smallskip\noindent\textbf{Realistic Rendering.}
Realistic rendering in games is gaining more consideration, as the research community is moving towards transferring the algorithms to real-world scenarios.
Some of such platforms are \cite{brodeur2017home,qiu2017unrealcv,yan2018chalet,kolve2017ai2,wu2018building}.
In Fig. \ref{fig-realistic}, we report an objective comparison of the most realistic scenes provided in these works against those in our platform. The results show that our platform provides a realistic rendering effect at the same level as the best of them.

\begin{figure*}
	\centering
	\subcaptionbox{\textit{Boomer} (Arena)}{
			\includegraphics[width=0.15\textwidth]{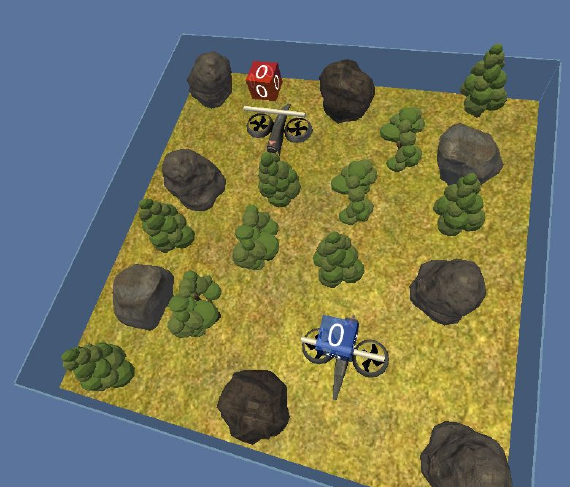}
	}
	\subcaptionbox{\textit{MsPacman} (Atari)}{
			\includegraphics[width=0.15\textwidth]{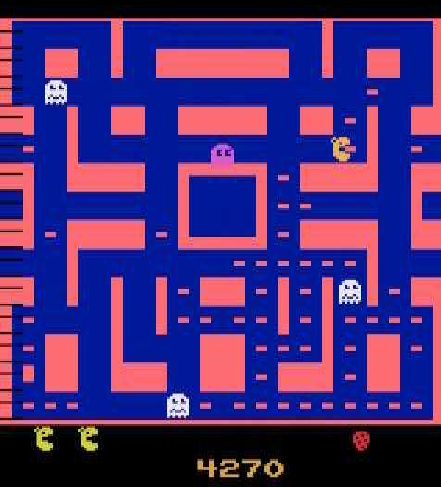}
	}
	\subcaptionbox{Comparison of train Frames Per Second (FPS) under different numbers of concurrent threads.}{
			\includegraphics[width=0.45\textwidth]{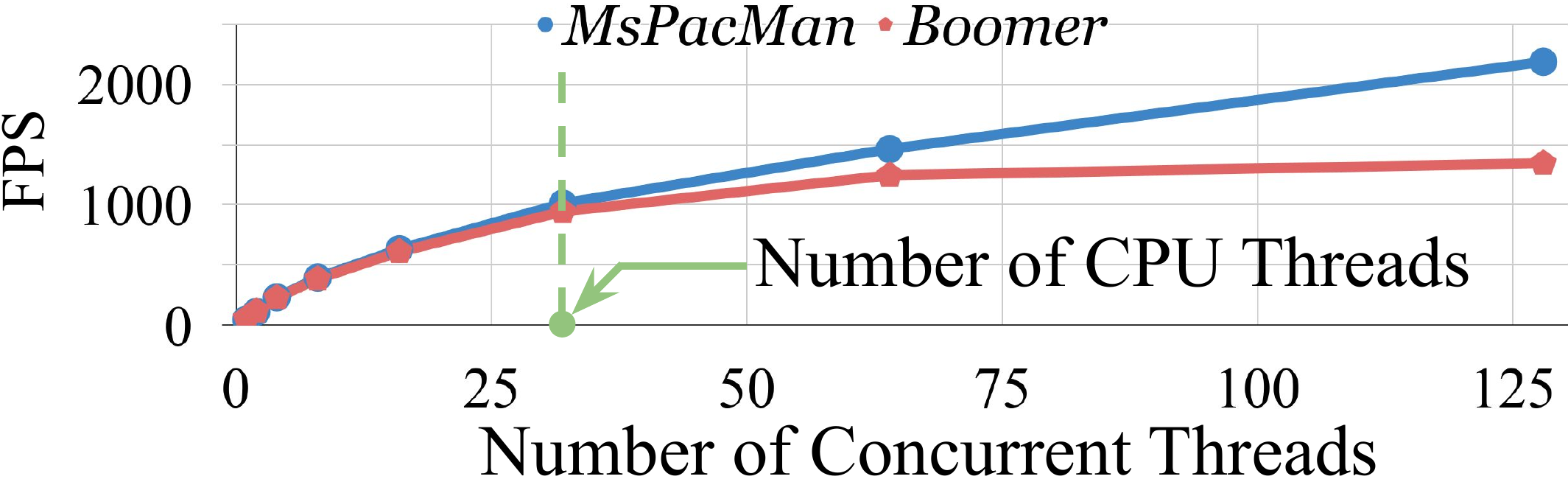}
	}
  \caption{Comparison of simulation speed.}
  \label{speed-comp}
\end{figure*}

\smallskip\noindent\textbf{Simulation Speed.}
Simulation speed and parallelism of an environment are two important factors for carrying out research.
Thus, we compare our game \textit{Boomer} in Fig. \ref{speed-comp} (a) with \textit{MsPacman} in Fig. \ref{speed-comp} (b) from the most widely used general evaluation platform ALE \cite{bellemare2013arcade}, both of which run on our parallel implementation of the PPO (D-PPO) baseline on a server with 32 CPU threads.
Then, we compare these two games, as they are of similar complexity. The result in Fig. \ref{speed-comp} (c) shows that Arena allows for parallel implementations while maintaining a similar simulation speed as ALE \cite{bellemare2013arcade} on games of similar complexity when the number of concurrent threads is below the number of CPU threads of the machine, i.e., smaller than 32.

\smallskip\noindent\textbf{Stochasticity.}
According to \cite{jaderberg2018human}, having enough stochasticity is essential for researchers to verify that their algorithms are learning general knowledge instead of memorizing action sequences.
Thus, we conduct a stochasticity study on the existing general evaluation platforms \textrm{ALE} \cite{bellemare2013arcade}, \textrm{Retro} \cite{nichol2018gotta}, \textrm{GVG-AI} \cite{perez2016general}, \textrm{Mujoco} \cite{todorov2012mujoco}, and \textrm{DM-Suite} \cite{tassa2018deepmind} by running a fixed sequence of 1000 actions repeatedly 1000 times and investigating how many branches are produced (averaged over all games in the corresponding platform).
Table \ref{table-stochasticity} shows that our platform generates most stochasticity among them, measured by the number of branches that the environment produces when applying the same action sequence on it.
This is accomplished by introducing stochasticity from the initial setup (e.g., a randomly generated map), the rendering effect (e.g., randomized light conditions and particle systems), and the physics system (e.g., randomized physics properties).

\section{How Verification for BMaRSs is Implemented}

Since the definitions of BMaRSs contain the statement of $\forall$, which requires enumerating over the entire space of some variables, we first set a hyper-parameter of $N$, denoting the number of samples that we will apply the verification operation to.
If a statement $\textbf{Y}$ of sample $i \in \mathcal{I}$ holds true for all $N$ samples from $\mathcal{I}$, we approximately make the judgement that $(\forall i \in \mathcal{I}, \textbf{Y})$ is true.
If a statement $\textbf{Y}$ of sample $i \in \mathcal{I}$ holds false for any one of $N$ samples from $\mathcal{I}$, we make the judgement that $(\forall i \in \mathcal{I}, \textbf{Y})$ is false.
Thus, in the following, we describe how to make the judgment of whether a statement $\textbf{Y}$ of sample $i$ holds true or not.

\textbf{Non-learnable} BMaRSs ($\mathcal{F}^{NL}$) are defined as:
\begin{equation}
\mathcal{F}^{NL} = \bigl\{
	f :
	\forall k \in \mathcal{K},
	\forall x \in \mathcal{X},
	\forall \pi_{x} \in \Pi_{x},
	\frac{\partial R^{f}_{x}}{\partial \pi_{x}} = \mathbf{0}
\bigr\},
\end{equation}
where $\mathbf{0}$ is a zero matrix of the same size and shape as the parameter space that defines $\pi_{x}$.
For each sample $i$, we first sample $k \in \mathcal{K},
x \in \mathcal{X},
\{\pi_{\hat{x}} \in \Pi_{\hat{x}}\}^{\hat{x} \in \mathcal{X}}$ randomly.
Then, we first run one episode to get $R^{f}_{x}$.
We denote it ${R^{f}_{x}}^{1}$, since it is from the first episode.
We run the second episode to get $R^{f}_{x}$ and denote it ${R^{f}_{x}}^{2}$, since it is from the second episode.
In the second episode, we keep the following variables the same as the first episode: $k$, $\{\pi_{\hat{x}}\}$ for all $ \hat{x} \in \mathcal{X} \setminus \{x\} $.
In the second episode, we sample a different $\pi_{x}$ from that in the first episode.
The $\pi_{x}$ in the first and second episode are denoted  $\pi_{x}^{1}$ and $\pi_{x}^{2}$ respectively.
Then, we can compute one sample of $\frac{\partial R^{f}_{x}}{\partial \pi_{x}} = \frac{{R^{f}_{x'}}^{1} - {R^{f}_{x'}}^{2}}{\pi_{x}^{1}-\pi_{x}^{2}}$ and make the judgement  whether $\frac{\partial R^{f}_{x}}{\partial \pi_{x}} = \mathbf{0}$ for sample $i$ holds true.

\textbf{Isolated} BMaRSs ($\mathcal{F}^{IS}$) are defined as:
\begin{equation}
	\mathcal{F}^{IS} = \bigl\{
		f :
		\forall k \in \mathcal{K},
		\forall x \in \mathcal{X},
		\forall x' \in \mathcal{X} \setminus \{x\},
		\forall \pi_{x} \in \Pi_{x},
		\forall \pi_{x'} \in \Pi_{x'},
		\frac{\partial R^{f}_{x}}{\partial \pi_{x'}} = \mathbf{0}
	\bigr\}.
	\end{equation}
For each sample $i$, we first sample $k \in \mathcal{K},
x \in \mathcal{X},
x' \in \mathcal{X} \setminus \{x\},
\{\pi_{\hat{x}} \in \Pi_{\hat{x}}\}^{\hat{x} \in \mathcal{X}}$ randomly.
Then, we first run one episode to get $R^{f}_{x}$.
We denote it  ${R^{f}_{x}}^{1}$, since it is from the first episode.
We run the second episode to get $R^{f}_{x}$ and denote it  ${R^{f}_{x}}^{2}$, since it is from the second episode.
In the second episode, we keep the following variables the same as the first episode: $k$, $\{\pi_{\hat{x}}\}$ for all $ \hat{x} \in \mathcal{X} \setminus \{x'\} $.
In the second episode, we sample a different $\pi_{x'}$ from that in the first episode.
The $\pi_{x'}$ in the first and second episode are denoted  $\pi_{x'}^{1}$ and $\pi_{x'}^{2}$ respectively.
Then, we can compute one sample of $\frac{\partial R^{f}_{x}}{\partial \pi_{x'}} = \frac{{R^{f}_{x}}^{1}-{R^{f}_{x}}^{2}}{\pi_{x'}^{1}-\pi_{x'}^{2}}$ and make the judgement whether $\frac{\partial R^{f}_{x}}{\partial \pi_{x'}} = \mathbf{0}$ for sample $i$ holds true.

\begin{table}
	\caption{Comparison of the stochasticity of platforms.}
	\begin{center}
		\resizebox{0.6\columnwidth}{!}{
			\begin{tabular}{cc*{9}{c}}
				& Platform
				& \textrm{ALE} & \textrm{Retro} & \textrm{GVG-AI} & \textrm{Mujoco} & \textrm{DM-Suite} & Arena
				\\
				\toprule
				& Branches
				& 0.0  & 431.2  & 23.1  & 12.2  & 50.2 & \textbf{922.4}
				\\
				\bottomrule
			\end{tabular}
		}
	\end{center}
	\label{table-stochasticity}
\end{table}

\textbf{Competitive} BMaRSs ($\mathcal{F}^{CP}$) are defined as:
\begin{equation}
\begin{split}
\mathcal{F}^{CP} = \bigl\{
	f :
		f \notin \mathcal{F}^{NL} \cup \mathcal{F}^{IS}
	\textrm{~and~}
		\forall k \in \mathcal{K},
		\forall x \in \mathcal{X},
		\forall \pi_{x} \in \Pi_{x},
		\forall \pi_{x'} \in \Pi_{x'}, \\
		\frac{\partial \int_{x' \in \mathcal{X}} R^{f}_{x'} dx'}{\partial \pi_{x}} = \mathbf{0}
\bigr\}.
\end{split}
\end{equation}
For each sample $i$, we first sample $k \in \mathcal{K},
x \in \mathcal{X},
\{\pi_{\hat{x}} \in \Pi_{\hat{x}}\}^{\hat{x} \in \mathcal{X}}$ randomly.
Then, we first run one episode to get $\int_{x' \in \mathcal{X}} R^{f}_{x'} dx'$.
We denote it  $\int_{x' \in \mathcal{X}} {R^{f}_{x'}}^{1} dx'$, since it is from the first episode.
We run the second episode to get $\int_{x' \in \mathcal{X}} R^{f}_{x'} dx'$ and denote it $\int_{x' \in \mathcal{X}} {R^{f}_{x'}}^{2} dx'$, since it is from the second episode.
In the second episode, we keep the following variables the same as the first episode: $k$, $\{\pi_{\hat{x}}\}$ for all $ \hat{x} \in \mathcal{X} \setminus \{x\} $ and we sample a different $\pi_{x}$ from that in the first episode.
The $\pi_{x}$ in the first and second episode are denoted  $\pi_{x}^{1}$ and $\pi_{x}^{2}$ respectively.
Then, we can compute one sample of $\frac{\partial \int_{x' \in \mathcal{X}} R^{f}_{x'} dx'}{\partial \pi_{x}} = \frac{\int_{x' \in \mathcal{X}} {R^{f}_{x'}}^{1} dx'-\int_{x' \in \mathcal{X}} {R^{f}_{x'}}^{2} dx'}{\pi_{x}^{1}-\pi_{x}^{2}}$ and make the judgement of whether $\frac{\partial \int_{x' \in \mathcal{X}} R^{f}_{x'} dx'}{\partial \pi_{x}} = \mathbf{0}$ for sample $i$ holds true.

\textbf{Collaborative} BMaRSs ($\mathcal{F}^{CL}$) are inspired by \cite{cai2011minmax} and defined as,
\begin{equation}
\begin{split}
\mathcal{F}^{CL} = \bigl\{
	f :
		f \notin \mathcal{F}^{NL} \cup \mathcal{F}^{IS}
	\textrm{~and~}
		\forall k \in \mathcal{K},
		\forall x \in \mathcal{X},
		\forall x' \in \mathcal{X} \setminus \{x\},
		\forall \pi_{x} \in \Pi_{x},
		\forall \pi_{x'} \in \Pi_{x'}, \\
		\frac{\partial R^{f}_{x'}}{\partial R^{f}_{x}} \geq 0
\bigr\}.
\end{split}
\end{equation}
For each sample $i$, we first sample $k \in \mathcal{K},
x \in \mathcal{X},
x' \in \mathcal{X} \setminus \{x\},
\{\pi_{\hat{x}} \in \Pi_{\hat{x}}\}^{\hat{x} \in \mathcal{X}}$ randomly.
Then, we first run one episode to get $R^{f}_{x}$ and $R^{f}_{x'}$.
We denote it  ${R^{f}_{x}}^{1}$ and ${R^{f}_{x'}}^{1}$, respectively, since it is from the first episode.
We run the second episode to get $R^{f}_{x}$ and $R^{f}_{x'}$ and denote it ${R^{f}_{x}}^{2}$ and ${R^{f}_{x'}}^{2}$, respectively, since it is from the second episode.
In the second episode, we keep the following variables the same as the first episode: $k$, $\{\pi_{\hat{x}}\}$ for all $ \hat{x} \in \mathcal{X} \setminus \{x\} $ and we sample a different $\pi_{x}$ from that in the first episode.
Then, we can compute one sample of $\frac{\partial R^{f}_{x'}}{\partial R^{f}_{x}} = \frac{{R^{f}_{x'}}^{1}-{R^{f}_{x'}}^{2}}{{R^{f}_{x}}^{1}-{R^{f}_{x}}^{2}}$ and make the judgement whether $\frac{\partial R^{f}_{x'}}{\partial R^{f}_{x}} \geq 0$ for sample $i$ holds true.

\textbf{Competitive and Collaborative Mixed} BMaRSs ($\mathcal{F}^{CC}$) are defined to be any situations other than the above four ones:
\begin{equation}
\label{mixed}
\mathcal{F}^{CC} = \bigl\{
	f :
	f \notin \mathcal{F}^{NL} \cup \mathcal{F}^{IS} \cup \mathcal{F}^{CP} \cup \mathcal{F}^{CL}
	\bigr\},
\end{equation}
which means if a reward function $f$ is verified to be not among above $\mathcal{F}^{NL}$, $\mathcal{F}^{IS}$, $\mathcal{F}^{CP}$ or $\mathcal{F}^{CL}$, it is verified to be among $\mathcal{F}^{CC}$.

\newtheorem{lemma}{Lemma}
\begin{lemma}
	\label{lemma-1}
	An equivalent expression of
	\begin{equation}
	\begin{split}
	\mathcal{F}^{CP} = \bigl\{
		f :
			f \notin \mathcal{F}^{NL} \cup \mathcal{F}^{IS}
		\textrm{~and~}
			\forall k \in \mathcal{K},
			\forall x \in \mathcal{X},
			\forall \pi_{x} \in \Pi_{x},
			\forall \pi_{x'} \in \Pi_{x'}, \\
			\frac{\partial \int_{x' \in \mathcal{X}} R^{f}_{x'} dx'}{\partial \pi_{x}} = \mathbf{0}
	\bigr\}
	\end{split}
	\end{equation}
	is:
	\begin{equation}
	\begin{split}
	\label{competitive}
	\mathcal{F}^{CP} = \bigl\{
	f :
		f \notin \mathcal{F}^{NL} \cup \mathcal{F}^{IS}
	\textrm{~and~}
		\forall k \in \mathcal{K},
		\forall x \in \mathcal{X},
		\forall \pi_{x} \in \Pi_{x},
		\forall \pi_{x'} \in \Pi_{x'}, \\
		\int_{x' \in \mathcal{X} } \frac{ \partial R^{f}_{x'}}{\partial R^{f}_{x}} dx' = 0
	\bigr\}.
	\end{split}
	\end{equation}
\end{lemma}

\noindent \textbf{Proof of Lemma~\ref{lemma-1}:}
Following the defination of $\mathcal{F}^{CP} $, since $f \notin \mathcal{F}^{NL}$, we have:
\begin{equation}
\exists k \in \mathcal{K},
\exists x \in \mathcal{X},
\exists \pi_{x} \in \Pi_{x},
\frac{\partial R^{f}_{x}}{\partial \pi_{x}} \neq \mathbf{0}.
\end{equation}
Thus, $\mathcal{F}^{CP} $ equals to:
\begin{equation}
\begin{split}
	\mathcal{F}^{CP} = \bigl\{
	f :
		f \notin \mathcal{F}^{NL} \cup \mathcal{F}^{IS}
	\textrm{~and~}
		\forall k \in \mathcal{K},
		\forall x \in \mathcal{X},
		\forall \pi_{x} \in \Pi_{x},
		\forall \pi_{x'} \in \Pi_{x'}, \\
		\int_{x' \in \mathcal{X} } \frac{ \partial R^{f}_{x'}}{\partial \pi_{x}} dx' \frac{\partial \pi_{x}}{\partial R^{f}_{x}}  = 0
	\bigr\}
\end{split}
\end{equation}
\begin{equation}
\begin{split}
\phantom{\mathcal{F}^{CP}} = \bigl\{
f :
	f \notin \mathcal{F}^{NL} \cup \mathcal{F}^{IS}
\textrm{~and~}
	\forall k \in \mathcal{K},
	\forall x \in \mathcal{X},
	\forall \pi_{x} \in \Pi_{x},
	\forall \pi_{x'} \in \Pi_{x'}, \\
	\int_{x' \in \mathcal{X} } \frac{ \partial R^{f}_{x'}}{\partial R^{f}_{x}} dx' = 0
\bigr\}, 
\end{split}
\end{equation}
completing the proof.

\begin{lemma}
	\label{lemma-2}
	\textit{Rock, Paper, and Scissors} in normal-form game \cite{myerson2013game} is a special case of $\mathcal{F}^{CP}$, which is defined in \eqref{competitive}.
\end{lemma}

\begin{table}
	\centering
	\setlength{\extrarowheight}{2pt}
	\begin{tabular}{ccc|c|c}
		& \multicolumn{1}{c}{} & \multicolumn{3}{c}{Player 2}\\
		& \multicolumn{1}{c}{} & \multicolumn{1}{c}{Strategy 1: Rock}  & \multicolumn{1}{c}{Strategy 2: Paper} & \multicolumn{1}{c}{Strategy 3: Scissors} \\
		\multirow{3}*{Player 1}
			& Strategy 1: Rock & $(0,0)$ & $(-1,1)$ & $(1,-1)$ \\\cline{3-5}
			& Strategy 2: Paper & $(1,-1)$ & $(0,0)$ & $(-1,1)$ \\\cline{3-5}
			& Strategy 3: Scissors & $(-1,1)$ & $(1,-1)$ & $(0,0)$ \\
	\end{tabular}
	\vspace{1em}
	\caption{Payoff matrix of \textit{Rock, Paper, and Scissors}.}
	\label{payoff-rps}
\end{table}

{\color{black}\noindent \textbf{Proof of Lemma~\ref{lemma-2}:}
The payoff matrix in a normal-form game \cite{myerson2013game} is a matrix aligning the tuple of $\{\textit{player},\textit{strategy}\}$ with the \textit{payoff} received by each player. The reward scheme in a multi-agent learning context is a map aligning the tuple of $\{\textit{agent}~x,\textit{policy}~\pi_{x}\}$ to the \textit{episode reward} $R^{f}_{x}$ received by the agent $x$.
Thus, \textit{player} is equivalent to \textit{agent}, \textit{strategy} is a special case of \textit{policy} when the episode length is one, \textit{payoff} is equivalent to \textit{episode reward} $R^{f}_{x}$ when the episode length is one.
Payoff matrix of \textit{Rock, Paper, and Scissors} is shown in Table \ref{payoff-rps}.
By selecting any pair of cells in the payoff matrix, say $(1,-1)$ and $(-1,1)$, we can compute a $\frac{\partial R^{f}_{x'} }{\partial R^{f}_{x}} = \frac{1-(-1)}{(-1)-1}$.
Thus, by enumerating over all possible pairs of cells in the payoff matrix to compute all $\frac{\partial R^{f}_{x'} }{\partial R^{f}_{x}}$, we can verify that the payoff matrix satisfies the definition of $\mathcal{F}^{CP}$ in \eqref{competitive}.}

\begin{lemma}
	\label{lemma-3}
	\textit{Symmetric zero-sum games} and \textit{cyclic games} in \cite{balduzzi2019open} are a special case of $\mathcal{F}^{CP}$, which is defined in \eqref{competitive}.
\end{lemma}

\noindent \textbf{Proof of Lemma~\ref{lemma-3}:}
In \cite{balduzzi2019open}, $W$ denotes a set of agents parameterized by the weights of a neural net.
For instance, $\textbf{w} \in W$ and $\textbf{v} \in W$ are two agents of specific policies from the full policy set $W$.
$\phi(\textbf{v}, \textbf{w})$ evaluates a pair of agents $(\textbf{w}, \textbf{v})$
\begin{equation}
	\phi: W \times W \rightarrow \mathbb{R}.
\end{equation}
The higher $\phi(\textbf{v}, \textbf{w})$, the better for agent $\textbf{v}$.
They refer to $\phi>0$, $\phi<0$, and $\phi=0$ as wins, losses, and ties, respectively, for $\textbf{v}$.
A game is defined as \textit{symmetric zero-sum game} if:
\begin{equation}
	\label{symmetric-zero-sum-game}
	\forall \textbf{w} \in W,
	\forall \textbf{v} \in W,
	\phi(\textbf{v}, \textbf{w}) = -\phi(\textbf{v}, \textbf{w}).
\end{equation}
A \textit{cyclic game} is a \textit{symmetric zero-sum game}.
As can be seen, when the game is symmetric
\begin{equation}
	\forall x \in \mathcal{X}, \Pi_{x}~\textrm{is the same and}~\Pi_{x}=\Pi_{\mathcal{X}},
\end{equation}
$W$ is equivalent to $\Pi_{\mathcal{X}}$.
When the game has only two players $x_{1}$ and $x_{2}$
\begin{equation}
	\label{two-player-game}
	\{x_{1}\} = \mathcal{X} \setminus \{x_{2}\},
\end{equation}
$\phi$ is equivalent to $R_{x_{1}}$.
Thus, $\textbf{v}$ and $\textbf{w}$ are equivalent to $\pi_{x_{1}}$ and $\pi_{x_{2}}$, respectively and the equivalent expression of \textit{symmetric zero-sum game} in  \eqref{symmetric-zero-sum-game} is:
\begin{equation}
	\label{symmetric-zero-sum-game-eq}
	\forall \pi_{x_{1}} \in \Pi_{\mathcal{X}},
	\forall \pi_{x_{2}} \in \Pi_{\mathcal{X}},
	R_{x_{1}} = -R_{x_{2}}.
\end{equation}
The equivalent expression of \eqref{competitive} when there are only two players $x_{1}$ $x_{2}$, i.e., satisfying \eqref{two-player-game}, is:
\begin{equation}
\begin{split}
\label{f-cp-two-player}
\mathcal{F}^{CP} = \bigl\{
f :
	f \notin \mathcal{F}^{NL} \cup \mathcal{F}^{IS}
\textrm{~and~}
	\forall x_{1} \in \mathcal{X},
	\{x_{2}\} = \mathcal{X} \setminus \{x_{1}\},
	\forall \pi_{x_{1}} \in \Pi_{\mathcal{X}},
	\forall \pi_{x_{2}} \in \Pi_{\mathcal{X}},\\
	\frac{ \partial R_{x_{1}}}{\partial R_{x_{2}}} = -1
\bigr\}.
\end{split}
\end{equation}
This shows that \textit{symmetric zero-sum games} are a special case of $\mathcal{F}^{CP}$.
Since \textit{cyclic games} are \textit{symmetric zero-sum games}, \textit{cyclic games} are also a special case of $\mathcal{F}^{CP}$.

\newcommand{\TableRaise}{-0.8 }

\newpage
\begin{landscape}
\begin{longtable}{| p{.183\linewidth} | p{.12\linewidth} |  p{.08\linewidth} |  p{.10\linewidth} |  p{.40\linewidth} |}
	\toprule
	Game & Name & Observation & Action Space & Description
	\\ \cmidrule(r){1-1}\cmidrule(lr){2-2}\cmidrule(lr){3-3}\cmidrule(lr){4-4}\cmidrule(lr){5-5}
	\raisebox{\TableRaise\totalheight}{\includegraphics[width=\WidthPicLongTable\textwidth, height=20mm]{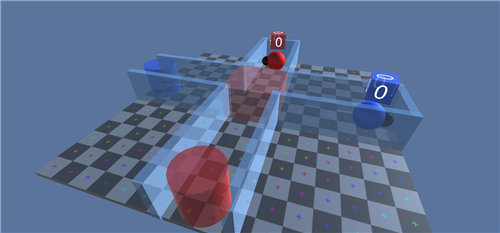}}
	&	\textit{Crossroads-2T1P-v1.}
	& Visual;
	& Continuous / Discrete;
	& Get to the target at the other side of the crossroad. The game uses dense reward function of IS\_Target at agent level (RewardSchemeScale at this level is 1), and use reward function of CP\_Ranking at global level (RewardSchemeScale at this level is 100).
	\\ \cmidrule(r){1-1}\cmidrule(lr){2-2}\cmidrule(lr){3-3}\cmidrule(lr){4-4}\cmidrule(lr){5-5}
	\raisebox{\TableRaise\totalheight}{\includegraphics[width=\WidthPicLongTable\textwidth, height=20mm]{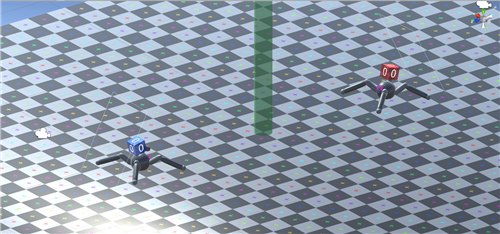}}
	&	\textit{ArenaCrawlerMove-2T1P-v1.}
	& Visual / RAM;
	& Continuous;
	& Get to the target in the middle. The game uses dense reward function of IS\_Target at agent level (RewardSchemeScale at this level is 1), specifically, it uses IsRewardMovingTowardsTarget and IsPenaltyHeadDown, and use reward function of CP\_Ranking at global level (RewardSchemeScale at this level is 100).
	\\ \cmidrule(r){1-1}\cmidrule(lr){2-2}\cmidrule(lr){3-3}\cmidrule(lr){4-4}\cmidrule(lr){5-5}
	\raisebox{\TableRaise\totalheight}{\includegraphics[width=\WidthPicLongTable\textwidth, height=20mm]{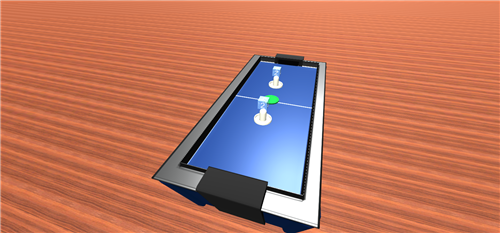}}
	&	\textit{AirHockey.}
	& Visual;
	& Continuous / Discrete;
	& See \href{https://en.wikipedia.org/wiki/Air_hockey}{{(\textcolor{brown}{\underline{\textit{Wiki Page}}})}}.


	\\ \cmidrule(r){1-1}\cmidrule(lr){2-2}\cmidrule(lr){3-3}\cmidrule(lr){4-4}\cmidrule(lr){5-5}
	\raisebox{\TableRaise\totalheight}{\includegraphics[width=\WidthPicLongTable\textwidth, height=20mm]{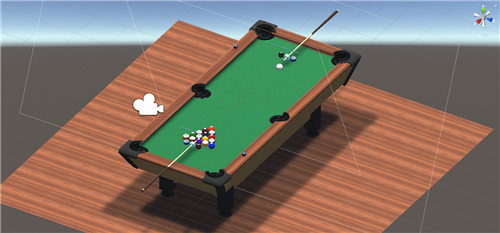}}
	& \textit{Billiards.}
	& Visual;
	& Discrete;
	& See \href{https://en.wikipedia.org/wiki/Chinese_eight-ball}{{(\textcolor{brown}{\underline{\textit{Wiki Page}}})}}.

	\\ \cmidrule(r){1-1}\cmidrule(lr){2-2}\cmidrule(lr){3-3}\cmidrule(lr){4-4}\cmidrule(lr){5-5}
	\raisebox{\TableRaise\totalheight}{\includegraphics[width=\WidthPicLongTable\textwidth, height=20mm]{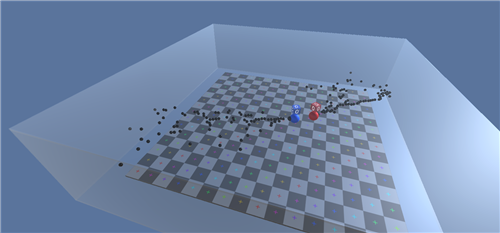}}
	& \textit{BlowBlow.}
	& Visual;
	& Continuous / Discrete;
	& Stay on the playground, push the opponent off the playground using your body, or the particles blown from you.
	\\ \cmidrule(r){1-1}\cmidrule(lr){2-2}\cmidrule(lr){3-3}\cmidrule(lr){4-4}\cmidrule(lr){5-5}
	\raisebox{\TableRaise\totalheight}{\includegraphics[width=\WidthPicLongTable\textwidth, height=20mm]{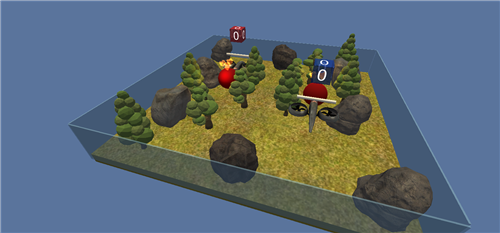}}
	& \textit{Boomer.}
	& Visual;
	& Discrete;
	& Inspired by \href{https://en.wikipedia.org/wiki/Bomberman_(1983_video_game)}{{(\textcolor{brown}{\underline{\textit{Wiki Page}}})}}.
	\\ \cmidrule(r){1-1}\cmidrule(lr){2-2}\cmidrule(lr){3-3}\cmidrule(lr){4-4}\cmidrule(lr){5-5}
	\raisebox{\TableRaise\totalheight}{\includegraphics[width=\WidthPicLongTable\textwidth, height=20mm]{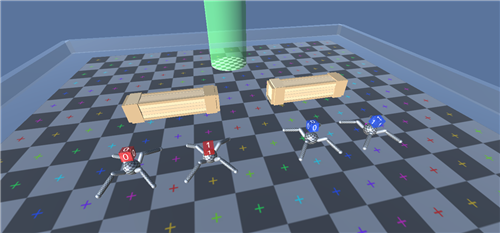}}
	& \textit{PushBox.}
	& Visual;
	& Continuous (forces at every joint of the robot ant);
	& Push the box forward as a team, and try to push the box to the target point first.
	\\ \cmidrule(r){1-1}\cmidrule(lr){2-2}\cmidrule(lr){3-3}\cmidrule(lr){4-4}\cmidrule(lr){5-5}
	\raisebox{\TableRaise\totalheight}{\includegraphics[width=\WidthPicLongTable\textwidth, height=20mm]{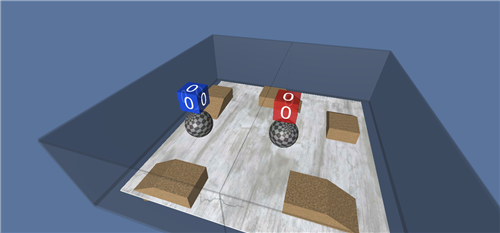}}
	&	\textit{OffTheGround.}
	& Visual;
	& Continuous / Discrete;
	& Stay on the playground, push the opponent using your body and try to take advantages of the terrain. \\
	\\ \cmidrule(r){1-1}\cmidrule(lr){2-2}\cmidrule(lr){3-3}\cmidrule(lr){4-4}\cmidrule(lr){5-5}
	\raisebox{\TableRaise\totalheight}{\includegraphics[width=\WidthPicLongTable\textwidth, height=20mm]{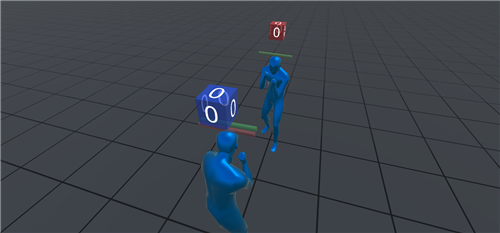}}
	&	\textit{KickBoxing.}
	& Visual;
	& Discrete;
	& Learn to use 18 different movements, such as Jab, Cross, Hook, Uppercut and etc.
		See \href{https://en.wikipedia.org/wiki/Kickboxing}{{(\textcolor{brown}{\underline{\textit{Wiki Page}}})}} for more information about the movements.
		Different movements cost different amount of energy.
		If the energy runs out, no movements can be made.
		Hitting different parts of the opponent's body results in different levels of harm.
		Thus, the agent is expected to learn to hit the soft part of the opponent's body with the hard part of its own body, meanwhile, try to keep a budget of the energy cost.
	\\ \cmidrule(r){1-1}\cmidrule(lr){2-2}\cmidrule(lr){3-3}\cmidrule(lr){4-4}\cmidrule(lr){5-5}
	\raisebox{\TableRaise\totalheight}{\includegraphics[width=\WidthPicLongTable\textwidth, height=20mm]{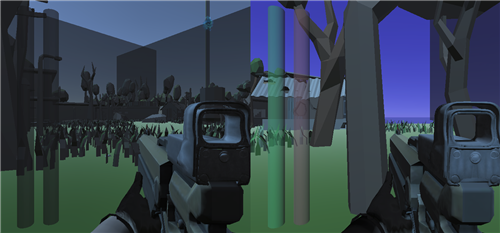}}
	&	\textit{MiniPUBG-Village.}
	& Visual;
	& Discrete;
	& \href{https://en.wikipedia.org/wiki/PlayerUnknown\%27s_Battlegrounds}{{(\textcolor{brown}{\underline{\textit{Wiki Page}}})}} for basic rules of \textit{PlayerUnknown's Battlegrounds} (PUBG).
		In this map of \textit{Village}, the agent is expected to learn to take various advantages of the buildings, grass and trees.
	\\ \cmidrule(r){1-1}\cmidrule(lr){2-2}\cmidrule(lr){3-3}\cmidrule(lr){4-4}\cmidrule(lr){5-5}
	\raisebox{\TableRaise\totalheight}{\includegraphics[width=\WidthPicLongTable\textwidth, height=20mm]{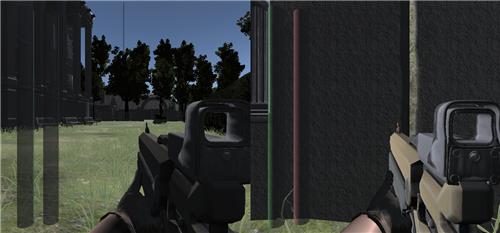}}
	& \textit{MiniPUBG-FloodedGround.}
	& Visual;
	& Discrete;
	& See \href{https://en.wikipedia.org/wiki/PlayerUnknown\%27s_Battlegrounds}{{(\textcolor{brown}{\underline{\textit{Wiki Page}}})}} for basic rules of \textit{PlayerUnknown's Battlegrounds} (PUBG).
	In this map of \textit{FloodedGround}, the playground is relatively large, so this game is designed for agents to learn in a larger population and learn long term planning.
	\\ \cmidrule(r){1-1}\cmidrule(lr){2-2}\cmidrule(lr){3-3}\cmidrule(lr){4-4}\cmidrule(lr){5-5}
	\raisebox{\TableRaise\totalheight}{\includegraphics[width=\WidthPicLongTable\textwidth, height=20mm]{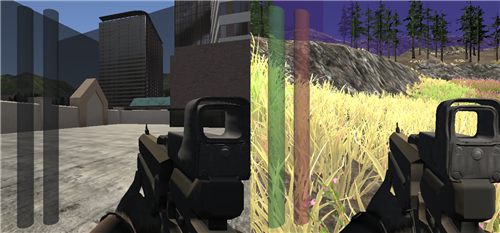}}
	&	\textit{MiniPUBG-Windridge.}
	& Visual;
	& Discrete;
	& See \href{https://en.wikipedia.org/wiki/PlayerUnknown\%27s_Battlegrounds}{{(\textcolor{brown}{\underline{\textit{Wiki Page}}})}} for basic rules of \textit{PlayerUnknown's Battlegrounds} (PUBG).
	In this map of \textit{FloodedGround}, there are both city areas and grass land areas.
	Thus, the agent is expected to learn the strategies of surviving between city and grassland areas.
	\\ \cmidrule(r){1-1}\cmidrule(lr){2-2}\cmidrule(lr){3-3}\cmidrule(lr){4-4}\cmidrule(lr){5-5}
	\raisebox{\TableRaise\totalheight}{\includegraphics[width=\WidthPicLongTable\textwidth, height=20mm]{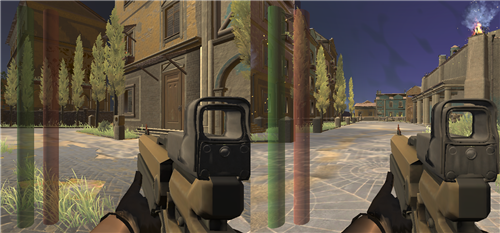}}
	&	\textit{MiniPUBG-SunTemple.}
	& Visual;
	& Discrete;
	& See \href{https://en.wikipedia.org/wiki/PlayerUnknown\%27s_Battlegrounds}{{(\textcolor{brown}{\underline{\textit{Wiki Page}}})}} for basic rules of \textit{PlayerUnknown's Battlegrounds} (PUBG).
		In this map of \textit{SunTemple}, there are mainly various buildings.
		Thus, the agent is expected to learn to hide, fight and take the advantages of the buildings.
	\\ \cmidrule(r){1-1}\cmidrule(lr){2-2}\cmidrule(lr){3-3}\cmidrule(lr){4-4}\cmidrule(lr){5-5}
	\raisebox{\TableRaise\totalheight}{\includegraphics[width=\WidthPicLongTable\textwidth, height=20mm]{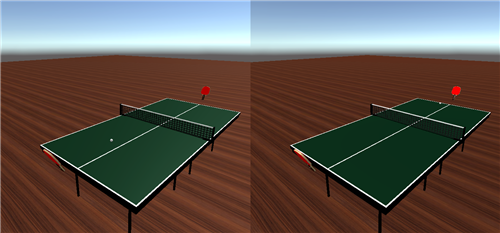}}
	&	\textit{PingPong.}
	& Visual;
	& Continuous / Discrete;
	& Adopt with full rules of standard table tennis, see \href{https://en.wikipedia.org/wiki/Table_tennis}{{(\textcolor{brown}{\underline{\textit{Wiki Page}}})}}.
	\\ \cmidrule(r){1-1}\cmidrule(lr){2-2}\cmidrule(lr){3-3}\cmidrule(lr){4-4}\cmidrule(lr){5-5}
	\raisebox{\TableRaise\totalheight}{\includegraphics[width=\WidthPicLongTable\textwidth, height=20mm]{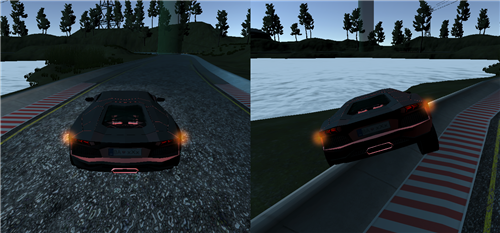}}
	&	\textit{RealRace-Lakes.}
	& Visual;
	& Discrete;
	& This track is general.
	\\ \cmidrule(r){1-1}\cmidrule(lr){2-2}\cmidrule(lr){3-3}\cmidrule(lr){4-4}\cmidrule(lr){5-5}
	\raisebox{\TableRaise\totalheight}{\includegraphics[width=\WidthPicLongTable\textwidth, height=20mm]{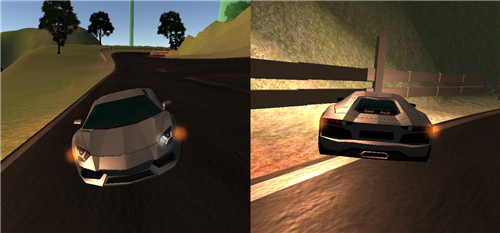}}
	&	\textit{RealRace-Sprint.}
	& Visual;
	& Discrete;
	& This track contains many long straight roads, where driving too fast will result in the car easily losing control.
		Thus, the agent is expected to learn to maintain the speed within a range that is high enough but safe and controllable at the same time.
	\\ \cmidrule(r){1-1}\cmidrule(lr){2-2}\cmidrule(lr){3-3}\cmidrule(lr){4-4}\cmidrule(lr){5-5}
	\raisebox{\TableRaise\totalheight}{\includegraphics[width=\WidthPicLongTable\textwidth, height=20mm]{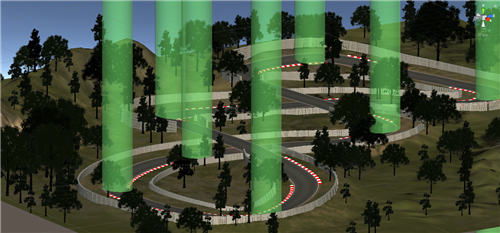}}
	& \textit{RealRace-Drift.}
	& Visual;
	& Discrete;
	& This track is designed for learning drafting.
	\\ \cmidrule(r){1-1}\cmidrule(lr){2-2}\cmidrule(lr){3-3}\cmidrule(lr){4-4}\cmidrule(lr){5-5}
	\raisebox{\TableRaise\totalheight}{\includegraphics[width=\WidthPicLongTable\textwidth, height=20mm]{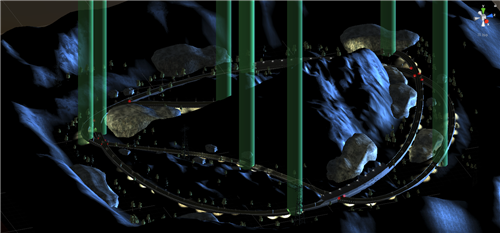}}
	&	\textit{RealRace-Night.}
	& Visual;
	& Discrete;
	& This scene is at night, so the agent is dealing with more complex night rendering conditions.
		Besides, this track contains lots of crossroads, where the agent needs to learn to recognize the road signs.
	\\ \cmidrule(r){1-1}\cmidrule(lr){2-2}\cmidrule(lr){3-3}\cmidrule(lr){4-4}\cmidrule(lr){5-5}
	\raisebox{\TableRaise\totalheight}{\includegraphics[width=\WidthPicLongTable\textwidth, height=20mm]{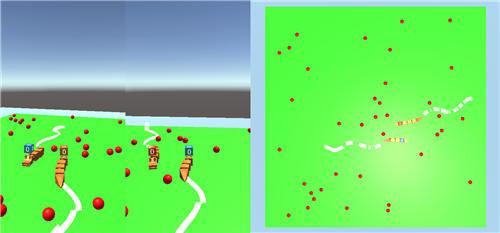}}
	&	\textit{Snake.}
	& Visual;
	& Continuous / Discrete;
	& Based on a classic video game, see \href{https://en.wikipedia.org/wiki/Snake_(video_game_genre)}{{(\textcolor{brown}{\underline{\textit{Wiki Page}}})}}.
		If two players collide with each other, the one of smaller length will be killed.
	\\ \cmidrule(r){1-1}\cmidrule(lr){2-2}\cmidrule(lr){3-3}\cmidrule(lr){4-4}\cmidrule(lr){5-5}
	\raisebox{\TableRaise\totalheight}{\includegraphics[width=\WidthPicLongTable\textwidth, height=20mm]{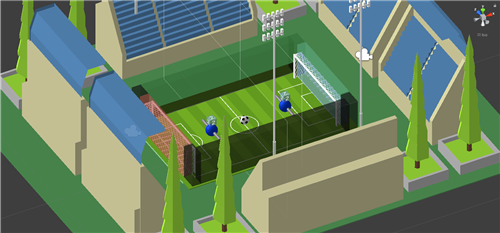}}
	& \textit{Soccers.}
	& Visual;
	& Continuous / Discrete;
	& Adopted from \cite{liu2019emergent}.
	\\ \cmidrule(r){1-1}\cmidrule(lr){2-2}\cmidrule(lr){3-3}\cmidrule(lr){4-4}\cmidrule(lr){5-5}
	\raisebox{\TableRaise\totalheight}{\includegraphics[width=\WidthPicLongTable\textwidth, height=20mm]{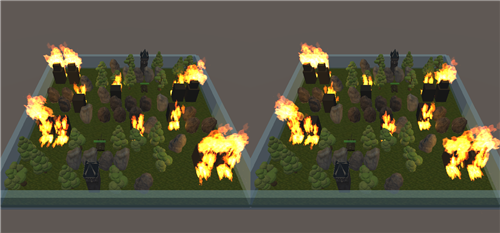}}
	&	\textit{BattleCity-TP.}
	& Visual;
	& Continuous / Discrete;
	& See \href{https://en.wikipedia.org/wiki/Battle_City_(video_game)}{{(\textcolor{brown}{\underline{\textit{Wiki Page}}})}}.
		This version of BettleCity is of third-person view.
	\\ \cmidrule(r){1-1}\cmidrule(lr){2-2}\cmidrule(lr){3-3}\cmidrule(lr){4-4}\cmidrule(lr){5-5}
	\raisebox{\TableRaise\totalheight}{\includegraphics[width=\WidthPicLongTable\textwidth, height=20mm]{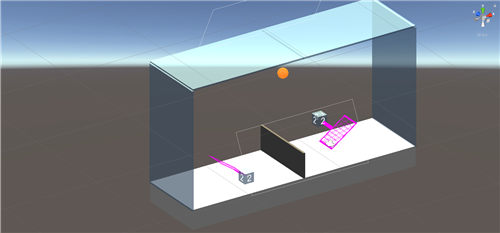}}
	& \textit{Tennis.}
	& Visual;
	& Continuous / Discrete;
	& Adopted from Unity ML-Agents \cite{juliani2018unity}.
	\\ \cmidrule(r){1-1}\cmidrule(lr){2-2}\cmidrule(lr){3-3}\cmidrule(lr){4-4}\cmidrule(lr){5-5}
	\raisebox{\TableRaise\totalheight}{\includegraphics[width=\WidthPicLongTable\textwidth, height=20mm]{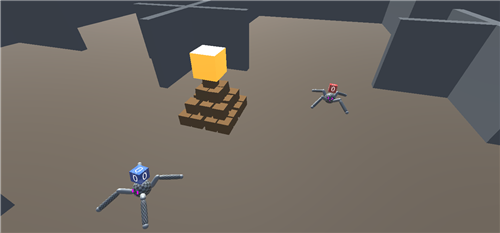}}
	&	\textit{DestroyPyRAMid.}
	& Visual / RAM;
	& Continuous;
	& The goal is to destroy the pyramid.
	\\ \cmidrule(r){1-1}\cmidrule(lr){2-2}\cmidrule(lr){3-3}\cmidrule(lr){4-4}\cmidrule(lr){5-5}
	\raisebox{\TableRaise\totalheight}{\includegraphics[width=\WidthPicLongTable\textwidth, height=20mm]{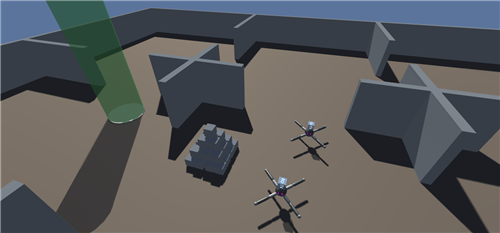}}
	&	\textit{TransportBricks.}
	& Visual / RAM;
	& Continuous;
	& The task is to transport all bricks to the target point.
	\\ \cmidrule(r){1-1}\cmidrule(lr){2-2}\cmidrule(lr){3-3}\cmidrule(lr){4-4}\cmidrule(lr){5-5}
	\raisebox{\TableRaise\totalheight}{\includegraphics[width=\WidthPicLongTable\textwidth, height=20mm]{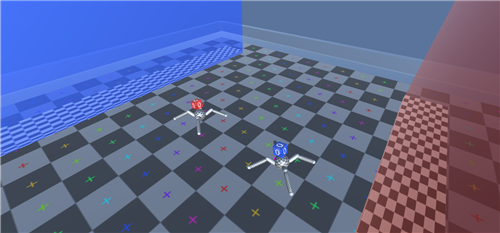}}
	&	\textit{RunToGoal.}
	& Visual / RAM;
	& Continuous;
	& Adopted from \cite{bansal2018emergent}.
	\\ \cmidrule(r){1-1}\cmidrule(lr){2-2}\cmidrule(lr){3-3}\cmidrule(lr){4-4}\cmidrule(lr){5-5}
	\raisebox{\TableRaise\totalheight}{\includegraphics[width=\WidthPicLongTable\textwidth, height=20mm]{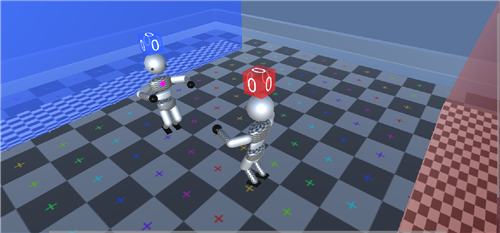}}
	&	\textit{YouShallNotPass.}
	& Visual / RAM;
	& Continuous;
	& Adopted from \cite{bansal2018emergent}.
	\\ \cmidrule(r){1-1}\cmidrule(lr){2-2}\cmidrule(lr){3-3}\cmidrule(lr){4-4}\cmidrule(lr){5-5}
	\raisebox{\TableRaise\totalheight}{\includegraphics[width=\WidthPicLongTable\textwidth, height=20mm]{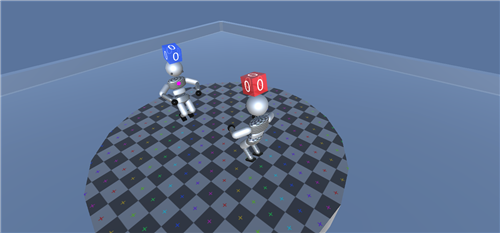}}
	&	\textit{Sumo.}
	& Visual / RAM;
	& Continuous;
	& Adopted from \cite{bansal2018emergent}.
	\\ \cmidrule(r){1-1}\cmidrule(lr){2-2}\cmidrule(lr){3-3}\cmidrule(lr){4-4}\cmidrule(lr){5-5}
	\raisebox{\TableRaise\totalheight}{\includegraphics[width=\WidthPicLongTable\textwidth, height=20mm]{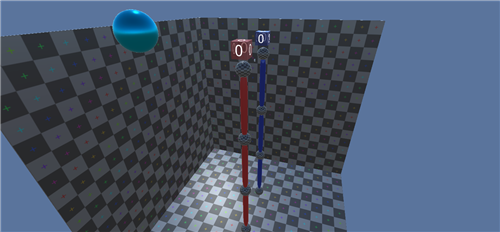}}
	&	\textit{Reacher.}
	& Visual / RAM;
	& Continuous;
	& Reach the target first.
	\\ \cmidrule(r){1-1}\cmidrule(lr){2-2}\cmidrule(lr){3-3}\cmidrule(lr){4-4}\cmidrule(lr){5-5}
	\raisebox{\TableRaise\totalheight}{\includegraphics[width=\WidthPicLongTable\textwidth, height=20mm]{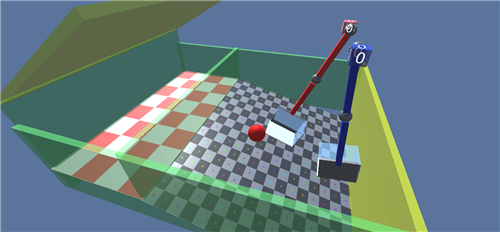}}
	&	\textit{LiftBall.}
	& Visual / RAM;
	& Continuous;
	& Collaborate to lift the ball.
	\\ \cmidrule(r){1-1}\cmidrule(lr){2-2}\cmidrule(lr){3-3}\cmidrule(lr){4-4}\cmidrule(lr){5-5}
	\raisebox{\TableRaise\totalheight}{\includegraphics[width=\WidthPicLongTable\textwidth, height=20mm]{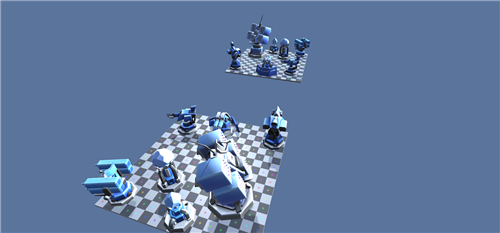}}
	&	\textit{TowerDefense.}
	& Visual;
	& Discrete;
	& Adopted from \cite{tian2017elf}.
	\\ \cmidrule(r){1-1}\cmidrule(lr){2-2}\cmidrule(lr){3-3}\cmidrule(lr){4-4}\cmidrule(lr){5-5}
	\raisebox{\TableRaise\totalheight}{\includegraphics[width=\WidthPicLongTable\textwidth, height=20mm]{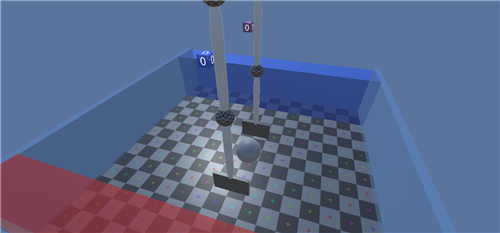}}
	&	\textit{PushBall.}
	& Visual / RAM;
	& Continuous;
	& Control the robot arm to push the ball to the base of the opponent.
	\\ \cmidrule(r){1-1}\cmidrule(lr){2-2}\cmidrule(lr){3-3}\cmidrule(lr){4-4}\cmidrule(lr){5-5}
	\raisebox{\TableRaise\totalheight}{\includegraphics[width=\WidthPicLongTable\textwidth, height=20mm]{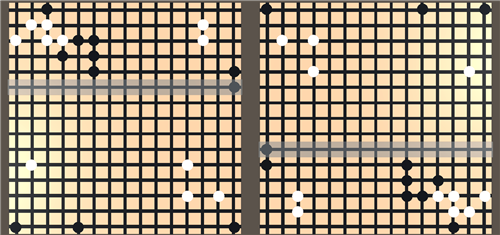}}
	&	\textit{GoBang.}
	& Visual / RAM;
	& Discrete;
	& See \href{https://en.wikipedia.org/wiki/Gomoku}{{(\textcolor{brown}{\underline{\textit{Wiki Page}}})}}.
	\\ \cmidrule(r){1-1}\cmidrule(lr){2-2}\cmidrule(lr){3-3}\cmidrule(lr){4-4}\cmidrule(lr){5-5}
	\raisebox{\TableRaise\totalheight}{\includegraphics[width=\WidthPicLongTable\textwidth, height=20mm]{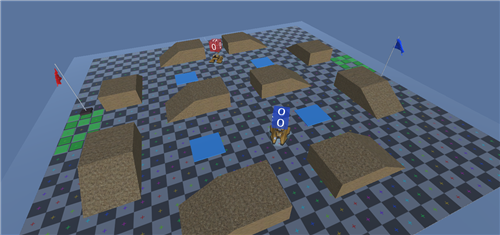}}
	&	\textit{CaptureFlag.}
	& Visual / RAM;
	& Discrete;
	& Adopted from \cite{jaderberg2018human}.
	\\ \cmidrule(r){1-1}\cmidrule(lr){2-2}\cmidrule(lr){3-3}\cmidrule(lr){4-4}\cmidrule(lr){5-5}
	\raisebox{\TableRaise\totalheight}{\includegraphics[width=\WidthPicLongTable\textwidth, height=20mm]{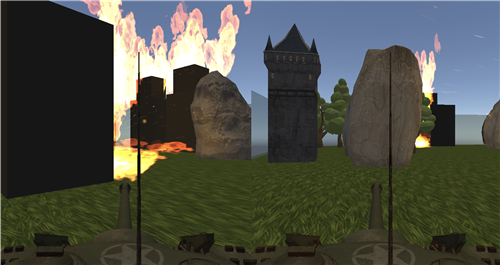}}
	&	\textit{BattleCity-FP.}
	& Visual;
	& Continuous / Discrete;
	& See \href{https://en.wikipedia.org/wiki/Battle_City_(video_game)}{{(\textcolor{brown}{\underline{\textit{Wiki Page}}})}}.
		This version of BettleCity is of first-person view.
	\\ \cmidrule(r){1-1}\cmidrule(lr){2-2}\cmidrule(lr){3-3}\cmidrule(lr){4-4}\cmidrule(lr){5-5}
	\raisebox{\TableRaise\totalheight}{\includegraphics[width=\WidthPicLongTable\textwidth, height=20mm]{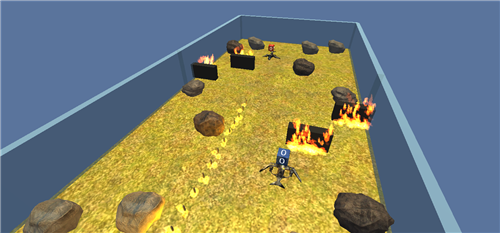}}
	&	\textit{Fighter.}
	& Visual;
	& Continuous / Discrete;
	& Take advantages of various bunkers and destroy the opponent.
	\\ \cmidrule(r){1-1}\cmidrule(lr){2-2}\cmidrule(lr){3-3}\cmidrule(lr){4-4}\cmidrule(lr){5-5}
	\raisebox{\TableRaise\totalheight}{\includegraphics[width=\WidthPicLongTable\textwidth, height=20mm]{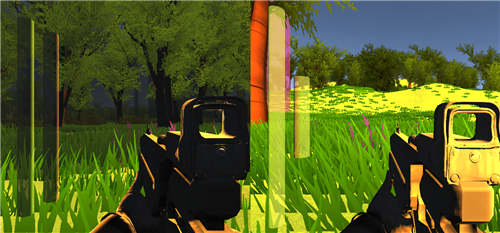}}
	&	\textit{MiniPUBG-Forest.}
	& Visual;
	& Discrete;
	& See \href{https://en.wikipedia.org/wiki/PlayerUnknown\%27s_Battlegrounds}{{(\textcolor{brown}{\underline{\textit{Wiki Page}}})}} for basic rules of \textit{PlayerUnknown's Battlegrounds} (PUBG).
	In this map of \textit{FloodedGround}, there are mainly grasslands.
	Thus, the agent is expected to learn the strategies of hiding and attacking.
	\\ \cmidrule(r){1-1}\cmidrule(lr){2-2}\cmidrule(lr){3-3}\cmidrule(lr){4-4}\cmidrule(lr){5-5}
	\raisebox{\TableRaise\totalheight}{\includegraphics[width=\WidthPicLongTable\textwidth, height=20mm]{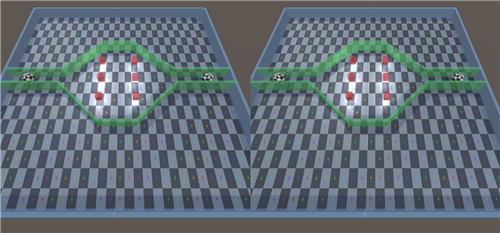}}
	&	\textit{FightForFood.}
	& Visual;
	& Continuous / Discrete;
	& A simple game where agents fight for limited food.
		When the food is exhausted, the episode ends.
	\\ \cmidrule(r){1-1}\cmidrule(lr){2-2}\cmidrule(lr){3-3}\cmidrule(lr){4-4}\cmidrule(lr){5-5}
	\raisebox{\TableRaise\totalheight}{\includegraphics[width=\WidthPicLongTable\textwidth, height=20mm]{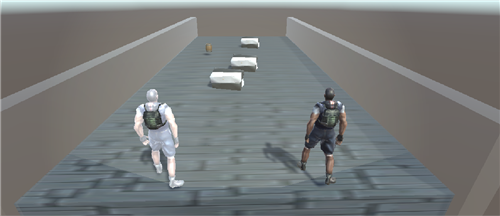}}
	& \textit{Runner.}
	& Visual;
	& Discrete;
	& A simple game where agents fight for higher score.
		When there is a score reach the specified threshold, the episode ends.
	\\ \bottomrule
\end{longtable}
\end{landscape}

\clearpage
\twocolumn

\bibliographystyle{aaai}
\bibliography{1513-yuhang}

\end{document}